\documentclass[preprint,showpacs,preprintnumbers,amsmath,amssymb]{revtex4}

\usepackage{graphicx}
\usepackage{epsfig}		
\usepackage{dcolumn}
\usepackage{bm}

\def\0{\mbox{\tiny $0$}}
\def\1{\mbox{\tiny $1$}}
\def\2{\mbox{\tiny $2$}}
\def\3{\mbox{\tiny $3$}}
\def\4{\mbox{\tiny $4$}}
\def\5{\mbox{\tiny $5$}}
\def\6{\mbox{\tiny $6$}}
\def\7{\mbox{\tiny $7$}}
\def\8{\mbox{\tiny $8$}}
\def\9{\mbox{\tiny $9$}}
\def\R{\mbox{\tiny $\mathit{R}$}}
\def\L{\mbox{\tiny $\mathit{L}$}}
\def\T{\mbox{\tiny $T$}}
\def\k{\mbox{\tiny $k$}}

\def\ii{\mbox{\tiny $i$}}
\def\I{\mbox{\tiny $I$}}
\def\D{\mbox{\tiny $D$}}
\def\A{\mbox{\tiny $\alpha$}}
\def\B{\mbox{\tiny $\beta$}}

\def\bb#1{\mbox{\footnotesize $(#1)$}}
\def\I{\mbox{\tiny $I$}}
\def\II{\mbox{\tiny $II$}}
\def\III{\mbox{\tiny $III$}}
\def\n{\mbox{\tiny $n$}}
\def\mi{\mbox{\tiny $-$}}
\def\pl{\mbox{\tiny $+$}}
\def\ppm{\mbox{\tiny $\pm$}}
\def\pmp{\mbox{\tiny $\mp$}}
\begin{document}

\title{Stationary phase method and delay times for relativistic and non-relativistic tunneling particles}

\author{A. E. Bernardini}
\affiliation{Departamento de F\'{\i}sica, Universidade Federal de S\~ao Carlos, PO Box 676, 13565-905, S\~ao Carlos, SP, Brasil}
\email{alexeb@ufscar.br}

\date{\today}



\pacs{02.30.Mv, 03.65.Xp, 03.65.Pm}






\begin{abstract}
This report deals with the basic concepts on deducing transit times for quantum scattering: the stationary phase method and its relation with delay times for relativistic and non-relativistic tunneling particles.
After reexamining the above-barrier diffusion problem, we notice that the applicability of this method is constrained by several subtleties in deriving the phase time that describes the localization of scattered wave packets.
Using a recently developed procedure - multiple wave packet decomposition - for some specifical colliding configurations, we demonstrate that the analytical difficulties arising when the stationary phase method is applied for obtaining phase (traversal) times are all overcome.
In this case, we also investigate the general relation between phase times and dwell times for quantum tunneling/scattering.
Considering a symmetrical collision of two identical wave packets with an one-dimensional barrier, we demonstrate that these two distinct transit time definitions are explicitly connected.
The traversal times are obtained for a symmetrized  (two identical bosons) and an antisymmetrized (two identical fermions) quantum colliding configuration.
Multiple wave packet decomposition shows us that the phase time (group delay) describes the exact position of the scattered particles and, in addition to the exact relation with the dwell time, leads to correct conceptual understanding of both transit time definitions.
At last, we extend the non-relativistic formalism to the solutions for the tunneling zone of a one-dimensional electrostatic potential in the relativistic (Dirac to Klein-Gordon) wave equation where the incoming wave packet exhibits the possibility of being almost totally transmitted through the potential barrier.
The conditions for the occurrence of accelerated and, eventually, superluminal tunneling transmission probabilities are all quantified and the problematic superluminal interpretation based on the non-relativistic tunneling dynamics is revisited.
Lessons concerning the dynamics of relativistic tunneling and the mathematical structure of its solutions suggest revealing insights into mathematically analogous condensed-matter experiments using electrostatic barriers in single- and bi-layer graphene, for which the accelerated tunneling effect deserves a more careful investigation.
\end{abstract}

\maketitle
\tableofcontents

\section{Introduction}

\hspace{1 em} The analytical methods utilized for reproducing the wave packet collision with a potential barrier have been widely discussed in the context of (one-dimensional) scattering and tunneling phenomena \cite{Hau89,Lan94,Per01}.
Within this extensively explored scenario, the stationary phase method (SPM) is the simplest and the most common approximation adopted for obtaining the group velocity of a wave packet in a quantum scattering process which represents the collision of a particle with a square (rectangular) barrier of potential.
First introduced to physics by Stokes and Kelvin \cite{K887}, the SPM essentially enables us to parameterize some subtleties of several quantum phenomena, such as tunneling \cite{Hau89,Ste93,Bro94}, resonances \cite{Con68,Bra70,Sok94B}, incidence-reflection transmission interferences \cite{Per01} as well as the Hartman effect \cite{Har62} and its superluminal traversal time interpretation \cite{Olk04,Lan94,Jak98}.
It is also often quoted in the context of testing different theories for temporal quantities such as arrival, dwell and delay times \cite{Hau89,Lan94,Win03} and the asymptotic behavior at long times \cite{Jak98,Bau01}.
The SPM is the simplest and most usual approximation method for describing the group velocity of a wave packet in a quantum scattering process represented by the collision of a particle with a potential barrier \cite{Olk04,Lan94,Hau87,Wig55,Har62,Ber04}.

To implement the accurate analysis with the SPM, it is necessary to work with the hypothesis of adding up the Fourier modes from the stationary solution to observe the evolution of a linear wave packet on collision with an obstacle.
Those interested in linear theory often look at stationary (frequency-domain) solutions of a single frequency, whereas those interested in nonlinear theory usually look at time-dependent pulses.
The time-dependent scattering of a wave-packet by a potential barrier in the linear Schr\"odinger equation is interesting because it reconciles two types of wave theories.
Their limitations, however, does not concern the domain of the method, i. e. its mathematical foundations and applicability, they focus on the indiscriminate use of the method for time-dependent scattering problems, in particular, for deriving tunneling phase times.

The issue of tunneling time is replete with controversy and paradoxes, preponderantly, because many tunneling time definitions seem to predict superluminal tunneling velocities.
Tunneling occurs when a wave impinges on a thin barrier of opaque material and some small amount of the wave {\em leaks} through to the other side.
In particular, the Hartman effect \cite{Har62} predicts that the tunneling time becomes independent of barrier length for thick enough barriers, ultimately resulting in superluminal tunneling velocities.
In this scenario, the correct quantification of the analytical incongruities which restrict the applicability of the principle of stationary phase is extended to the study of tunneling where, eventually, the superluminal interpretation of transition times can be ruined.

Trying to understand the Nature of the superluminal barrier tunneling has brought up a fruitful discussion in the literature \cite{But03,Win03A,Win03,Olk04} since pulses of light and microwaves appear to tunnel through a barrier at speeds faster than a reference pulse moves through a vacuum \cite{Nim92,Ste93,Spi94,Nim94,Hay01}.
The experiments that promoted the controversial discussions were performed with a lattice of layers of transparent and opaque materials arranged so that waves of some frequencies are reflected (through destructive interference) but other frequencies pass through the lattices in a kind of {\em filter} effect correlated with the Hartman effect \cite{Har62}.

To find a definitive answer for the time spent by particle to penetrate a classically forbidden region delimited by a potential barrier, people have introduced quantities that have the dimension of time and can somehow be associated with the passage of the particle through the barrier, i. e. the tunneling times.
These proposals have led to the introduction of several theoretical definitions \cite{Olk04,Baz67,But83,Hau87,Fer90,Yuc92,Hag92,Bro94,Sok94,Olk95,Jak98,Olk02}, some of which can be organized into three groups.
(1) The first group comprises a time-dependent description in terms of wave packets where some features of an incident packet and the comparable features of the transmitted packet are utilized to describe a quantifiable {\em delay} as a tunneling time \cite{Hau89}.
(2) In the second group the tunneling times are computed based on averages over a set of kinematical paths, whose distribution is supposed to describe the particle motion inside a barrier.
In this case, Feynman paths are used like real paths to calculate an average tunneling time with the weighting function $\exp{[i\, S\, x(t)/\hbar]}$, where $S$ is the action associated with the path $x(t)$ (where $x(t)$ represents the Feynman paths initiated from a point on the left of the barrier and ending at another point on the right of it \cite{Sok87}).
The Wigner distribution paths \cite{Bro94}, and the Bohm approach \cite{Ima97,Abo00} are included in this group.
(3) In the third group we notice the introduction of a new degree of freedom, constituting a physical clock for the measurements of tunneling times.
This group comprises the methods with a Larmor clock \cite{But83} or an oscillating barrier \cite{But82}.
Separately, standing on itself is the {\em dwell} time defined by the interval during which the incident flux has to exist and act, to provide the expected accumulated particle storage, inside the barrier \cite{Lan94}.

There is no general agreement \cite{Olk04,Olk92} among the above definitions about the meaning of tunneling times (some of the proposed tunneling times are actually traversal times, while others seem to represent in reality only the spread of their distributions) and about which, if any, of them is the proper tunneling time \cite{Olk04}.
In the context on which we intend to elaborate, the tunneling mechanism is embedded by theoretical constructions involving analytically-continuous {\em Gaussian}, or infinite-bandwidth step pulses to examine the tunneling process.
Nevertheless, such holomorphic functions do not have a well-defined front in a manner that the interpretation of the wave packet speed of propagation becomes ambiguous.
Moreover, infinite-bandwidth signals cannot propagate through any real physical medium (whose transfer function is therefore finite) without pulse distortion, which also leads to ambiguities in determining the propagation velocity during the tunneling process.

In what concerns such ambiguities, some of the barrier traversal time definitions lead, under tunneling time conditions, to very short times, which can even become negative.
It can precipitately induces an interpretation of violation of simple concepts of causality.
Otherwise, negative speeds do not seem to create problems with causality, since they were predicted both within special relativity and within quantum mechanics \cite{Olk95}.
A possible explanation of the time advancements related to the negative speeds can come, in any case, from consideration of the very rapid spreading of the initial and transmitted wave packets for large momentum distribution widths.
Due to the similarities between tunneling (quantum) packets and evanescent (classical) waves, exactly the same phenomena are to be expected in the case of classical barriers \footnote{In particular, we could mention the analogy between the stationary Helmholtz equation for an electromagnetic wave packet - in a waveguide, for instance - in the presence of a {\em classical} barrier and the stationary Schr\"odinger equation, in the presence of a potential barrier \cite{Lan94,Jak98,Nim94}).}.
The existence of such negative times is predicted by relativity itself based on its ordinary postulates \cite{Olk04}, and they appear to have been experimentally detected in many works \cite{Gar70,Chu82}.

In all cases described by the non-relativistic (Schr\"odinger) dynamics \cite{Olk04}, the pulse (wave packet) that emerges from the tunneling process is greatly attenuated and front-loaded due to the {\em filter} effect (only the leading edge of the incident wave packet survives the tunneling process without being severally attenuated to the point that it cannot be detected).
If one measures the speed by the peak of the pulse, it looks faster than the incident wave packet.
Since the transmission probability depends analytically on the momentum component $k$  ($T \equiv T\bb{k}$) the initial (incident wave) momentum distribution can be completely distorted by the presence of the barrier of potential.
As there is no sharp beginning to a pulse, we cannot declare the instant of its arrival at a certain point.
Thus the computation of the tunneling time becomes fundamentally meaningless.
We could only watch the rising edge of the pulse and try to recognize what is arriving.

Meanwhile, in some limiting cases of the relativistic (Klein) tunneling phenomena where the relativistic kinetic energy is approximately equal to the potential energy of the barrier, and $m c L /\hbar << 1$, particles with mass $m$ can pass through a potential barrier $V_{\0}$ of width $L$ with transmission probability $T$ approximately equal to one.
Differently from the non-relativistic tunneling analysis, the original momentum is kept undistorted and there is no {\em filter} effect.
Our final aim is thus to demonstrate that the tunneling time is then computed for a completely undistorted transmitted wave packet, which legitimizes any eventual accelerated transmission.

In this scenario, the first part of our study is to characterize some limitations on the use of the SPM for obtaining the right position of a propagating wave packet subject to the most common one-dimensional potential configuration.
In section II, the analysis of the applicability of the SPM is concentrated on the investigation of the non-relativistic diffusion of an incoming single wave packet with energy spectrum totally localized above the potential barrier.
We are particularly interested in obtaining an analytical description of the above-barrier diffusion problem by pointing out some interpretive problems which concern with non-conservation of probability during the collision process.
By recurring to a wave packet multiple peak decomposition \cite{Ber04}, we can decompose the original colliding wave packet into {\em multiple} reflected and transmitted components by means of which the conservation of probabilities is recovered during the scattering process.
The analytic expressions for all the reflected, transmitted and intermediary components are obtained and the validity of them is discussed in terms of an illustrative example where the analytic approximation and the reflects exactly the numerical results.

In section III, we present the theoretical constructions of the physical framework where the study of a tunneling process is embedded.
It involves analytically-continuous {\em Gaussian} pulses which due to its not well-defined front lead to ambiguities in the interpretation of the wave packet speed of propagation.
Beside it, infinite bandwidth signals cannot propagate through any real physical medium (whose transfer function is therefore finite) without pulse distortion, which also leads to ambiguities in determining propagation velocity during the tunneling process.
Moreover, some of the barrier transposing time definitions lead, in tunneling time conditions, to very short times which can even become negative where they may seem to contradict simple concepts of causality.

To partially overcome these incompatibilities, in section IV, we discuss a theoretical exercise involving a symmetrical collision with a one-dimensional square potential where we recompose the scattered wave packet by summing the amplitudes of the reflected and transmitted waves in the scope of what we classify as a multiple peak decomposition analysis \cite{Ber04}, where the conditions for the SPM applicability are totally recovered.

The general and explicit relation between phase times and dwell times for quantum tunneling or scattering is investigated in section V.
For the quoted symmetrical colliding configuration, the phase time gives the exact position of the {\em center of mass} of each symmetrically scattered wave packet.
Consequently, we can have a realistic idea of the {\em magnitude} the dwell time for the case in which the energy of the incident particle is smaller than the barrier potential energy (tunneling configuration).
In spite of a theoretical focus, the results here obtained apply only to such configurations which should deserve further attention by experimenter, while the existing experiments report the inaccurate results on non-symmetrical configurations.

In section VI we compute the tunneling transmission probabilities and the respective delay-times for wave packets when a relativistic dynamic is taken into account \cite{Ber08I,Ber08II}.
The transmission probabilities, the phase times and the dwell times for the relativistic dynamics are all obtained and the conditions for the occurrence of accelerated and, eventually, superluminal tunneling transmission probabilities are specified.

We draw our conclusions in section VII pointing out suggestive insights into condensed-matter experiments using electrostatic barriers in single- and bi-layer graphene, for which the accelerated tunneling effect deserves a more careful investigation.

\section{The stationary phase method in the scattering problem}

The principle of stationary phase to describe the group velocity is well-established, and was employed by Sommerfeld and Brillouin in their early studies of wave propagation and group velocity \cite{Bri60}.
The method has in time become a standard tool for several theoretical applications \cite{PBE} adopted not only by physicists but also by biologists, economists etc.
It can also be used in a statistical sense, such that the most likely events tend to be associated with slowly-varying phase variation in the frequency domain, and unlikely  events tend to be associated with rapidly varying phase with frequency.

In analyzing mathematical problems in physics, one often finds it desirable to know the behavior of a function for large values of some parameter $s$, that is, the asymptotic behavior of the function.
Specific examples are furnished by the well-known {\em Gamma} function and various {\em Bessel} functions \cite{Arf01}.
All these analytic functions are defined by integrals
\small\begin{equation}
I(s)  = \int_{_C}^{^{}}dz \, H(z, s),
\label{100}
\end{equation}\normalsize
where $H$ is analytic in $z$ and depends on a real parameter $s$.
In the most cases of physical interest the integrals like $I(s)$ can be redefined in terms of
\small\begin{equation}
H(z, s)  = h(z) \, \exp{\left[s \, f(z)\right]},
\label{101}
\end{equation}\normalsize
which leads to the explicit resolution of $I(s)$, for large values of $s$, by an asymptotic approximation procedure named {\em steepest descent} method (SDM).
The key idea in this method is that as $s$ grows very large, the main contribution to the integral will come from values of $z$ very close to the points at which $df(z)/dz = 0$.
Irrespective of the nature of the particular function, for some paths of integration along which the imaginary part of $f(z)$ is constant, these points are called {\em saddle-points} because they are the points where the real and imaginary parts of $f(z)$ are stationary with respect to the position in the complex plane $(x,y)$ without being an absolute maximum or minimum.
In fact, if we have $f(z) = u(x,y) + i \, v(x,y)$, we can easily show from Cauchy-Riemann relations that along any path $v(x,y) =  constant$ the rate of change of $u(x,y)$ can only vanish at a {\em saddle-point}
\cite{Arf01}.
From any point, the direction in which $u(x,y)$ decreases most rapidly is one along which $v(x,y)$ is constant, and in this sense such paths are paths of {\em steepest descent}.
Perhaps the most common situation where the SDM is applicable is that in which the path of integration $v(x,y) =  constant$ runs from a {\em saddle-point} $z_{s}$ to infinite, with $u(x,y)$ decreasing monotonically all the way.
In a simplified analysis, the function $f(z)$ can be expanded around $z_{s}$ into a Taylor expansion.
If we replace $f(z)$ with the first two terms of this expansion in the exponential of Eq.~(\ref{101}),
\small\begin{equation}
f(z) \approx f(z_{s}) - \mu^{\2}(z)
\label{111},
\end{equation}\normalsize
with
\small\begin{equation}
\mu(z) = \sqrt{-\frac{f^{\prime\prime}(z_{s})}{2}}(z - z_{s}), \nonumber
\end{equation}\normalsize
and extend the limits of integration from $-\infty$ to $+ \infty$, we find
\small\begin{equation}
I(s)  = - \sqrt{2} f^{\prime\prime}(z_{s}) \exp{\left[s \, f(z_{s})\right]}\, \int_{_{-\infty}}^{^{+\infty}}
d\mu \, h(z(\mu))\, \exp{\left[ - s \, \mu^{\2}(z)\right]},
\label{112}
\end{equation}\normalsize
In extensions of this method complex analysis is used to find a contour of {\em steepest descent} for an equivalent integral expressed as a path integral.
To get the first term in the asymptotic expansion of (\ref{112}) the value of $\mu(z) \, h(z) / f^{\prime}(z)$ at $z = z_{s}$ is required.
Provided $f^{\prime\prime}(z_{s})$ is not zero and $h(z_{s})$ is not infinite the integral $I(s)$ results
\small\begin{equation}
I(s)  = \sqrt{-\frac{2 \,\pi}{s \, f^{\prime\prime}(z_{s})}} \, h(z_{s}) \, \exp{\left[s \, f(z_{s})\right]},
\label{113}
\end{equation}\normalsize
which, however, is {\em not} completely free of some particular supplementary conditions for its applicability \cite{Arf01}.
An alternative, and very similar, method to the SDM is the method of {\em stationary phase} (SPM).
Though perhaps less general and less immediately convincing analytically, it often has the advantage of closer contact with the physical problem.
The integral to be considered is more suitably written as
\small\begin{equation}
I(s)  = \int_{_C}^{^{}}dz \, h(z) \, \exp{[i\, s \, f(z)]},
\label{114}
\end{equation}\normalsize
where, in practice, the exponential commonly represents a traveling wave.
The SPM was first introduced explicitly by Lord Kelvin \cite{K887}.
A rigorous mathematical treatment which would justify the statements made above was subsequently done by Watson \cite{Wat18} and the complete discussion involving the asymptotic approximation was done by Focke \cite{Foc54}.
The deductions from the SPM follow much the same pattern as those from the SDM.
The assumed paths of integration are $v(x,y) =  constant$, which means that the amplitude part of the exponential is constant along the path, while the phase part varies most rapidly, a reversal of the situation in the SDM.
For asymptotic values of $s$ the rapid oscillations of $s \, f(z)$ over most of the range of integration means that the integrand averages to almost zero.
Exceptions to this cancellation rule occur only when the stationary phase condition is satisfied, i. e. the only significant contribution to the integral arises from portions of the path in the vicinity of {\em saddle-points} or {\em end-points}, but the physical interpretation of the mechanism by which this comes about is now in terms of phase interference rather than amplitude decay.
Thus the approximation corresponding to (\ref{113}) for the SDM is
\small\begin{equation}
I(s)  = \sqrt{-\frac{2 \,\pi}{s \, f^{\prime\prime}(z_{s})}} \, h(z_{s}) \, \mbox{$\exp{\left[\mi \frac{i \,\pi}{4}\right]}$}\exp{\left[i \,s \, f(z_{s})\right]},
\label{115}
\end{equation}\normalsize
for the SPM.
But one aspect in which there is some distinction between the methods should be noted.
With a {\em steepest descent} path which starts at a {\em saddle-point} and does not go to infinity, the contribution of the point at the end of the path to the strict asymptotic expansion is zero in comparison with the {\em saddle-point} contribution by the virtue of the extra exponential factor it contains.
On the other hand, with a stationary phase path of the same type the contribution of the point at the end of the path is, in general, of the order of that of the {\em saddle-point} merely divided by $\sqrt{s}$; it is excluded, therefore, from the asymptotic approximation only if the first term of the asymptotic expansion is alone retained.
To sum up, the methods of steepest descent and stationary phase, strip off their mathematical expression, depend on choosing a path of integration in such a way that the integrand, due to its exponential factor, contributes irrelevantly to the integral except in the vicinity of certain {\em saddle-points} (or {\em end-points}).

To illustrate the applicability of the SPM into a physical problem it should be instructive to assume an equivalence between the complex variable $z$ and a real one $k$ so that an integral which represents a free quantum particle propagation can be identified by the wave packet solution of the one-dimensional Schr\"odinger equation,
\small\begin{eqnarray}
i\hbar\frac{\partial}{\partial t} \psi(x, t)  &=& \left[-\frac{\hbar^{\2}}{2 m} \frac{d^{\2}}{dx^{\2}} + V\bb{x}\right] \psi(x, t),
\label{201b}
\end{eqnarray}\normalsize
with $V\bb{x} = 0$, described in terms of the integral,
\small\begin{eqnarray}
\psi(x, t)  &=& \int_{_{-\infty}}^{^{+\infty}}\frac{d k}{\sqrt{2\pi}}\,
G(k , k_{\0}) \, \exp{\left[- i\, (E\bb{k} \, t +  k \, x - k \, x_{\0})\right]},
\label{201}
\end{eqnarray}\normalsize
where we have set $\hbar = 1$ and the dispersion relation as $E\bb{k} = k^{\2}/(2m)$.
The function $G(k , k_{\0})$ physically represents a narrow momentum distribution centered around the momentum $k_{\0}$, and the Eq.~(\ref{201}) can be identified as the integral of Eq.~(\ref{114}) by simply rewriting $G(k , k_{\0})$ as $\sqrt{2\pi} \, h(z)$ and $s \, f(z)$ as the phase $i\,(k\, x - k \, x_{\0} - E \, t)$ in an extension to the complex plane ($k \rightarrow z$).

The integral (\ref{201}) can therefore be estimated by finding the value for which the phase has a vanishing derivative, evaluating (approximately) the integral in the neighborhood of this point.
The movement of the peak coordinate of the wave packet $\psi(x,t)$ can be obtained by imposing the stationary phase condition
\small\begin{equation}
\left.\frac{d}{dk}[E \, t - \, k (x - x_{\0})]\right|_{k = k_{s}} = 0
~~~~\Rightarrow~~~~x_{\mbox{\tiny max}} = x_{\0} + \frac{k_{s}}{m} \, t,
\label{202}
\end{equation}\normalsize
when $k_{s}=k_{\0}$, the maximum of $G(k, k_{\0})$, which means that the peak of the wave packet propagates with a velocity $v = \frac{k_{\0}}{m}$.
In fact, we ratify this result when we explicitly calculate the integral of
Eq.~(\ref{201}) by introducing a {\em Gaussian} momentum distribution given by
\small\begin{equation}
G(k , k_{\0}) = g(k - k_{\0}) = \left(\frac{a^{\2}}{2 \, \pi}\right)^{\frac{1}{4}}
\exp{\left[-\frac{a^{\2} (k -k_{\0})^{\2}}{4}\right]}.
\label{203}
\end{equation}\normalsize
In this case, the result of the integration (\ref{201}) gives us
\small\begin{equation}
\psi(x,t) = \varphi[x - x_{\0}, t],
\label{204A}
\end{equation}\normalsize
where
\small\begin{equation}
\varphi[x, t] = \left[ \mbox{$\frac{\,\,\,\pi a^{\2}}{2}$} \,
\left( 1 + \mbox{$\frac{4\,t^{\2}}{m^{\2}a^{\4}}$} \right)
\right]^{-\frac{1}{4}} \exp \left[ -
\frac{\left(x-\frac{k_{\0}}{m}\,t\right)^{\2}}{a^{\2}\left(
1+\frac{2\,i\,t}{ma^{\2}}\right)}-\mbox{$\frac{i}{2}$} \arctan
\mbox{$\left(\frac{2t}{ma^{\2}}\right)$}+i (k_{\0}x - E_{\0}t) \right],
\label{204}
\end{equation}\normalsize
which evidently confirms the result of Eq.~(\ref{202}).

Meanwhile, the method leads to some new interpretive discussion when the momentum distribution becomes a complex function, i. e. when $G(k , k_{\0})$ can be written as $|G(k, k_{\0})|\, \exp{[i\,\lambda(k)]}$.
In this case, the stationary phase condition for a free wave packet represented by Eq.~(\ref{202}) will be modified by the presence of an additional phase $\lambda(k)$.
By assuming that $\lambda(k)$ can be expanded in the neighborhood of $k_{\0}$, i. e.
\small\begin{equation}
\lambda(k) \approx \lambda(k_{\0}) + (k - k_{\0})
\left.\frac{d \lambda(k)}{dk}\right|_{k = k_{\0}},
\label{205}
\end{equation}\normalsize
the new stationary phase condition can hereby lead to
\small\begin{equation}
x_{\mbox{\tiny max}} = x_{\0} - \left.\frac{d \lambda(k)}{dk}\right|_{k = k_{\0}}
+ \frac{k_{\0}}{m} \, t,
\label{206}
\end{equation}\normalsize
which can be interpreted as space-time translation.
As it is commonly presented in several textbooks of quantum mechanics \cite{Coh77}, one can be persuaded to take the stationary phase condition as a necessary and {\em sufficient} statement for applying the method.
Nevertheless, although it can be used to predict the most likely outcome of an experiment, even when it is related to the applicability of the stationary phase principle, it does not exclude the possibility of alternative outcomes.
It is well-known in the literature that both SDM and SPM are applicable only if certain conditions apply.
As we shall demonstrate in the following section, the results of the SPM, when applied to the one-dimensional scattering problem, depend critically upon the manipulation of the generic amplitude $G(k,k_{\0})$ prior to the application of the method.

\subsection{The above-barrier diffusion problem}

\begin{figure}
\vspace{-0.8cm}
\begin{center}
\epsfig{file= 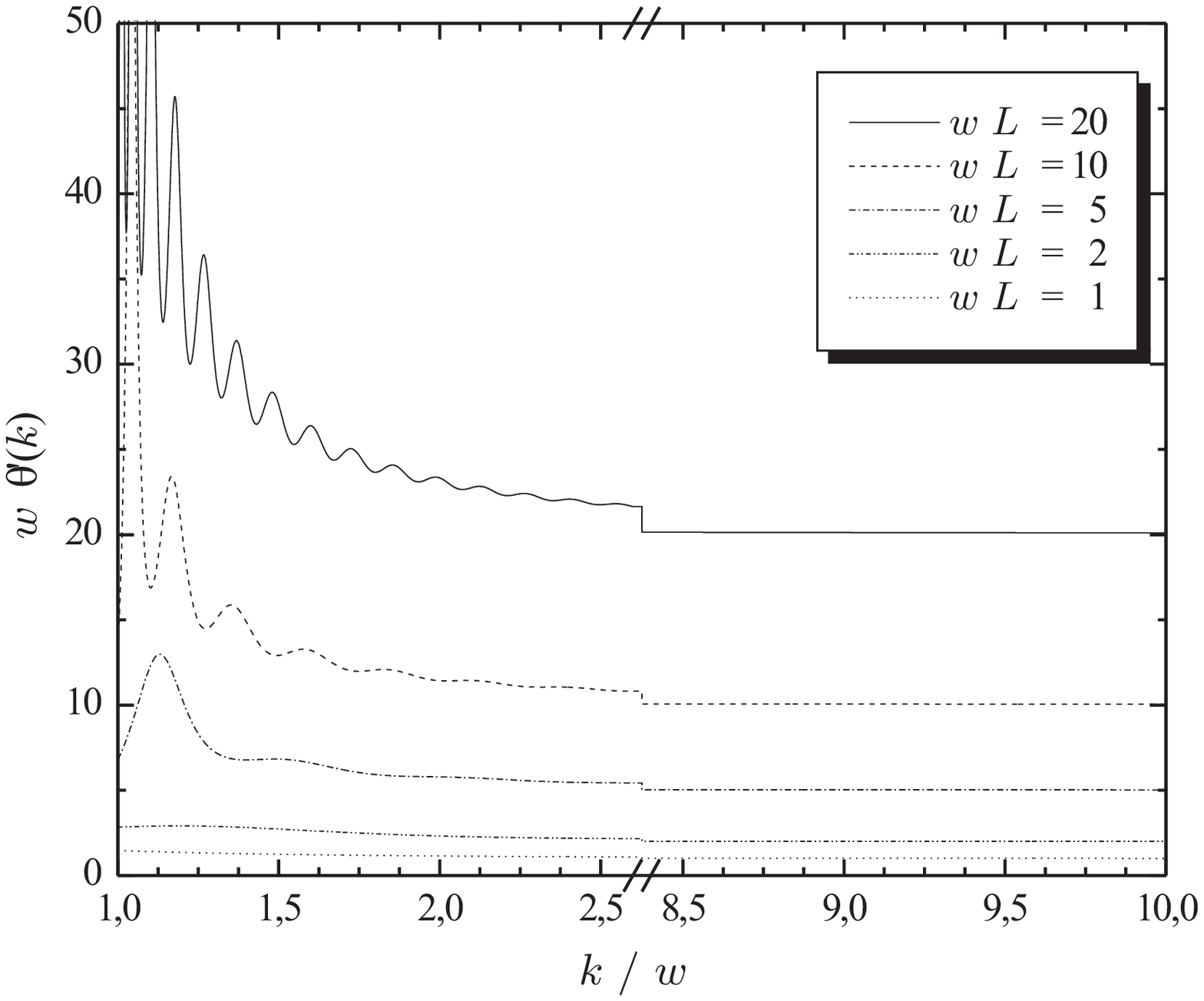, width = 10 cm}
\end{center}
\vspace{-0.8cm}
\caption{\footnotesize
The phase derivative $\Theta^{\prime}(k)$ dependence on $k/w$.
$\Theta^{\prime}(k)$ does not present an adequate analytical behavior (smoothness) for the applicability of the SPM when  $k$ approximates to $w$ since the phase derivative oscillates too rapidly (The phase is not stationary).
The method can be accurately applied for larger values of $k/w$ when the phase is really stationary.}
\label{fig0A}
\end{figure}

To elucidate some questions upon the ambiguities which appear when we utilize the SPM, let us study the scattering of an incoming wave packet by a potential barrier for propagating energies larger than the barrier upper limit $V_{\0}$.
The stationary wave solution of the Schr\"odinger equation (\ref{201b}) \footnote{The same analysis can be applied to relativistic wave equations, for instance, to the Dirac, Klein-Gordon and Maxwell equations.} obtained when we consider the potential barrier described by
\small\begin{equation}
\begin{array}{clll}  0&&~~~\mbox{if}~~~x < 0&~~~~\mbox{region I}, \\
 V_o &&~~~\mbox{if}~~~0 < x < L&~~~~\mbox{region II}, \\
  0 &&~~~\mbox{if}~~~x > L&~~~~\mbox{region III},
\end{array}
\label{p2}
\end{equation}\normalsize
can be decomposed into different wave functions for each interval of $x$, i. e.
\small\begin{equation}
\begin{array}{ccccccc}
\Phi(k,x) &=& \phi_{\I}(k, x) &  + &
                \phi_{\II}(k, x) & +  &\hspace*{.8cm}
                \phi_{\III}(k, x), \\
& & \overbrace{\phi_{Inc}(k, x) + \phi_{\R}(k, x)} &  &
                \overbrace{\phi_{\A}(k,x) +\phi_{\B}(k, x)} &  &
                 \hspace*{.8cm} \overbrace{\phi_{\T}(k, x)}
\end{array},
\label{p3}
\end{equation}\normalsize
where
\small\begin{eqnarray}
\phi_{Inc}(k, x) &=& \exp{\left[ i \,k \,x\right]},\nonumber\\
\phi_{\R}(k, x)  &=& R(k)\exp{\left[ - i \,k \,x\right]},\nonumber\\
\phi_{\A}(k, x)  &=& \alpha(k)\exp{\left[ i \,q  \,x\right]},\nonumber\\
\phi_{\B}(k, x)  &=& \beta(k)\exp{\left[ - i \,q  \,x\right]},\nonumber\\
\phi_{\T}(k, x)  &=& T(k)\exp{ \left[ i \,k \,x\right]},
\label{p4}
\end{eqnarray}\normalsize
with $q = (k^{\2} - w^{\2})^{\frac{1}{2}}$ and $w = \sqrt{2\, m \, V_{\0}}$.
By solving the constraint equations \cite{Coh77} which set the continuity conditions at $x = 0$ and $x = L$, we obtain
\small\begin{eqnarray}
\alpha(k) &=&     \left[\frac{k(k + q)}{\mathcal{F}(k)}\right]
\exp{\left[i\,\Theta(k) - i\,q \, L\right]},
~~~~~~\beta(k)  = -   \left[\frac{k(k - q)}{\mathcal{F}(k)}\right]
\exp{\left[i\,\Theta(k) + i\,q \, L\right]},\nonumber\\
R(k)      &=&   -i \,  \left[\frac{k^{\2} - q^{\2}}{\mathcal{F}(k)}\right]\sin{[q \,L]}
\exp{\left[i\,\Theta(k)\right]},
~~~~~~T(k)      =     \left[\frac{2 \, k \, q }{\mathcal{F}(k)}\right]
\exp{\left[i\,\Theta(k) - i\,k\, L\right]},
\label{p5}
\end{eqnarray}\normalsize
where
\small\begin{eqnarray}
\mathcal{F}(k) &=& \left\{4 \, k^{\2} \, q^{\2} \cos^{\2}{[q \,L]} +
\left(k^{\2} + q^{\2} \right)^{\2}\sin^{\2}{[q \,L]}\right\}^{\frac{1}{2}}
\end{eqnarray}\normalsize
and
\small\begin{eqnarray}
\Theta(k) &=& \arctan{\left\{\frac{k^{\2} + q^{\2} }{2 \, k \,q }\tan{[q \,L]}\right\}}.
\label{p6}
\end{eqnarray}\normalsize
The explicit expression for the correspondent propagating wave packets can be obtained by solving integrals like
\small\begin{eqnarray}
\psi_f(x, t)  &=& \int_{_{w}}^{^{+\infty}}\frac{d k}{\sqrt{2\pi}}\,
g(k - k_{\0}) \, \phi_f(k,x)\,\exp{[- i\,E \, t]},
\label{p6A}
\end{eqnarray}\normalsize
with $f \equiv \alpha,\,\beta,\, R,\,T$.
As a first approximation which is commonly used in quantum mechanics textbooks \cite{Coh77}, we obtain the analytical formulas to these integrals by assuming the momentum distribution $g(k - k_{\0})$ is sufficiently sharply peaked around the maximum point $k_{\0}$, with $k_{\0} > w$.
In this case, the integration can be  extended from $[w, \infty]$ to $[-\infty,\infty]$ without modifying the final result.
However, a careful investigation of $\Theta(k)$ in Eq.~(\ref{p6}) would clearly indicate that the above integral is not dominated by the exponential for certain values of $k_{\0}/w$ (see the Fig.~(\ref{fig0A} where we describe the derivative of $\Theta(k)$ as a function of $k_{\0}/w$).

\begin{figure}
\vspace{-0.8cm}
\begin{center}
\epsfig{file= 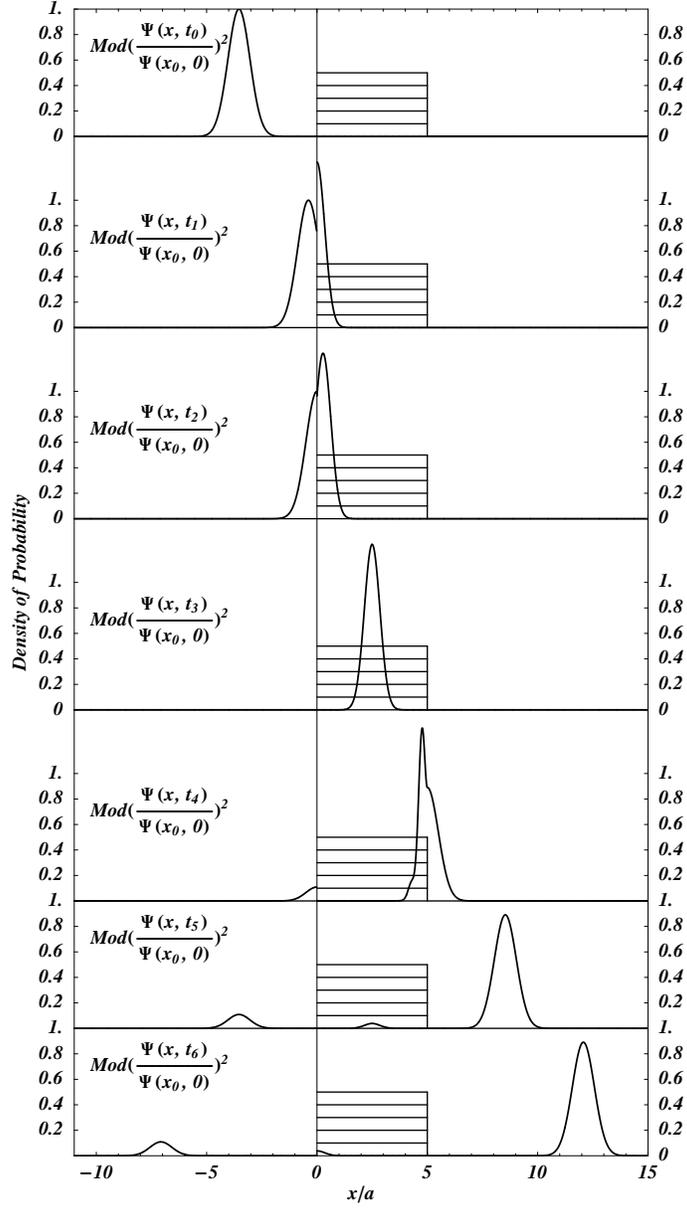, width = 9 cm}
\end{center}
\vspace{-0.8cm}
\caption{\footnotesize Erroneous interpretation of the scattering of an incoming wave packet by a one-dimensional potential
barrier of height $V_{\0}$ and width $L$.
It has the purpose of illustrating the deficiencies inherent to the {\em wrong} applicability of the SPM.
We have plotted the propagating wave packets at the corresponding times $t_{n} = (m a^{\2})[n\,(L/a)]/(a\, q_{\0})$ (with $n = 0,\,1,\,...,\,5$ and with the normalization constraint $m a^{\2} = 1$) by assuming the incoming wave
packet starts at $x =- (k_{\0}\,L)/(2q_{\0})$.
From the {\em false} behavior of the density of probabilities it becomes obvious that the total probability is not conserved as it was expected.
The square of the amplitude modulus would supposedly represent a collision of a wave packet of average width $a$
with a potential barrier $V_{\0}$ of width $L = 5\,a$ where, for illustrating reasons, we have adopted $k_{\0} = \sqrt{2}w$ and $w a = 10000$.
Only a fixed region in $x$ close to the barrier is shown.}
\label{fig1A}
\end{figure}

In the sense we are investigating, as we shall demonstrate in the following, it is {\em erroneously} assumed that the $k$-dependent phase terms $\Theta(k)$ and $q\equiv q(k)$ can be approximated by a series expansion around $k = k_{\0}$ up to the first order term, i. e.
\small\begin{equation}
\Theta(k) \approx \Theta(k_{\0}) + (k - k_{\0})\Theta^{\prime}(k_{\0})
~~~~\mbox{and}~~~~
q  \approx q_{\0}  + (k - k_{\0})\,
\left.\frac{d q }{d k}\right|_{k = k_{\0}},
\label{p6B}
\end{equation}\normalsize
where
\small\begin{equation}
\Theta^{\prime}(k)
 = \frac{2\,m}{q}
\left[\frac{\left(k^{\2} + q^{\2} \right)k^{\2} \,q \,L - \left(k^{\2} - q^{\2} \right)^{\2}
\sin{[q \,L]}\cos{[q \,L]}}{4 \, k^{\2} \, q^{\2}
+ \left(k^{\2} - q^{\2} \right)^{\2}\sin^{\2}{[q \,L]}}\right]
~~~~\mbox{and}~~~~
\left.\frac{d q }{d k}\right|_{k = k_{\0}} =
 \frac{k_{\0}}{q_{\0}}.
\label{p10}
\nonumber\\
\end{equation}\normalsize
At the same time, when $k$ is approximated by $k_{\0}$, it is assumed that the modulating amplitude $|\phi_f(k_{\0},x)|$ can be put out of the integral giving the following results,
\footnotesize\begin{eqnarray}
\psi_{Inc}(x, t) &\approx& \varphi[x - x_{\0}, t],\nonumber\\
\psi_{\R}(x, t)  &\approx& R(k_{\0})\,
\int_{_{-\infty}}^{^{+\infty}}\frac{d k}{\sqrt{2\pi}}\,
\mbox{$g(k - k_{\0}) \, \exp{\left[- i\,E \, t - i \,k (x + x_{\0})
+ i\, (k - k_{\0})\Theta^{\prime}(k_{\0})
\right]}$}\nonumber\\
& = & \mbox{$R(k_{\0})
\exp{\left[- i \, k_{\0}\,\Theta^{\prime}(k_{\0})\right]}\,
\varphi\left[- x - x_{\0} + \Theta^{\prime}(k_{\0}), t\right]$},\nonumber\\
\psi_{\A}(x, t)  &\approx& \alpha(k_{\0})\,
\int_{_{-\infty}}^{^{+\infty}}\frac{d k}{\sqrt{2\pi}}\,
\mbox{$g(k - k_{\0}) \, \exp{\left[- i \left(E \, t + k \, x_{\0}- q_{\0}\,x\right)
 + i \,(k - k_{\0})\left(\frac{k_{\0}}{q_{\0}}(x- L)
 + \Theta^{\prime}(k_{\0})\right)
\right]}$}\nonumber\\
& = & \mbox{$\alpha(k_{\0})
\exp{\left[i \, q_{\0} \, x - i \, \frac{k^2_{\0}}{q_{\0}}\,(x - L)
 - i \, k_{\0}\Theta^{\prime}(k_{\0})\right]}\,
\varphi\left[\frac{k_{\0}}{q_{\0}}(x - L)
  - x_{\0} + \Theta^{\prime}(k_{\0}), t\right]$},\nonumber\\
\psi_{\B}(x, t)  &\approx& \beta(k_{\0})
\int_{_{-\infty}}^{^{+\infty}}\frac{d k}{\sqrt{2\pi}}\,
\mbox{$g(k - k_{\0}) \, \exp{\left[- i \left(E \, t + k \, x_{\0}+ q_{\0}\,x\right)
- i \, (k - k_{\0})\left(\frac{k_{\0}}{q_{\0}}(x - L)
- \Theta^{\prime}(k_{\0})\right)
\right]}$}\nonumber\\
& = & \mbox{$\beta(k_{\0})
\exp{\left[- i \, q_{\0} \, x + i \, \frac{k^2_{\0}}{q_{\0}}\,(x - L)
 - i \, k_{\0}\Theta^{\prime}(k_{\0})\right]}\,
\varphi\left[-\frac{k_{\0}}{q_{\0}}(x - L)- x_{\0}
 + \Theta^{\prime}(k_{\0}), t\right]$},\nonumber\\
\psi_{\T}(x, t)  &\approx& T(k_{\0})\,
\int_{_{-\infty}}^{^{+\infty}}\frac{d k}{\sqrt{2\pi}}\,
\mbox{$g(k - k_{\0}) \, \exp{\left[- i\,E \, t + i \,k (x - x_{\0}) +
 i\, (k - k_{\0})\left(\Theta^{\prime}(k_{\0})- L\right)
\right]}$}\nonumber\\
& = & \mbox{$ T(k_{\0})
\exp{\left[- i \, k_{\0}\Theta^{\prime}(k_{\0})\right]}\,
\varphi\left[x - x_{\0} - L + \Theta^{\prime}(k_{\0}), t\right]$}.\nonumber\\
\label{p6C}
\end{eqnarray}\normalsize
By observing the {\em Gaussian} shape of $\varphi[x, t]$ given by Eq.~(\ref{204}), one could easily (but wrongly!) identify the position of the peak of the wave packets.
The times corresponding to the position $x$ of the incident and reflected wave packet peaks would be respectively given by
\small\begin{equation}
t_{Inc}(x) = \left.\left[\frac{x - x_{\0}}{v_k}\right]\right|_{k = k_{\0}} ~~~~\mbox{and}~~~~
t_{\R}(x) = \left.\left[-\frac{x + x_{\0} - \Theta^{\prime}(k)}{v_k}\right]\right|_{k = k_{\0}}
\label{p9}
\end{equation}\normalsize
where $v_k = \frac{d E}{d k} = \frac{k}{m}$.
Only $x < 0$ is physical in this result since these waves, by definition, lie in region I.
Since the phase of the incoming wave contains only the plane wave factors, i.e. it is devoid of $\Theta(k)$, the incoming peak reaches the barrier at $x=0$ at time $t_{Inc}(0) = -(x_{\0}/v_{k_{\0}})$ (neglecting interference effects).
The presence of the phase term $\Theta^{\prime}(k_{\0}) = (d \Theta(k)/dk)|_{k = k_{\0}}$ for $t_{\R}(x)$ {\em would} imply a time delay, or a time advance, which depends on the sign of $\Theta^{\prime}(k_{\0})$, for the reflected wave with respect to the incident one, i.e.
\small\begin{equation}
\left.t_{\R}(0) = t_{Inc}(0) -\frac{\Theta^{\prime}(k)}{v_k}\right|_{k = k_{\0}}
\label{p10A}
\end{equation}\normalsize
It is analogous to what happens for the step potential tunneling  penetration when $E < V_{\0}$ \cite{Coh77}.
When $0 < x < L$ in region II, the times for the ``intermediary'' $\alpha$ and $\beta$ wave packet peaks would be given by
\small\begin{equation}
t_{\A}(x) =\left.\left[ \frac{(x - L)}{v_q}
 - \frac{(x_{\0} - \Theta^{\prime}(k))}{v_k}\right]\right|_{k = k_{\0}}~~~~\mbox{and}~~~~
t_{\B}(x)  = \left.\left[ -\frac{(x - L)}{v_q}
-  \frac{(x_{\0} - \Theta^{\prime}(k))}{v_k}\right]\right|_{k = k_o}
\label{p12}
\end{equation}\normalsize
And finally, when $x > L$ in region III, the peak of the transmitted wave packet would be written as
\small\begin{equation}
t_{\T}(x) =\left.\left[ \frac{x - x_{\0} - L + \Theta^{\prime}(k)}{v_k} \right]\right|_{k = k_{\0}}
\label{p14}
\end{equation}\normalsize
The above results (\ref{p9}-\ref{p14}) could be directly obtained by applying the SPM to the wave functions $\psi_f(x, t)$ expressed in Eq.~(\ref{p6A}), where the stationary phases for each decomposed wave component $f$ would be given by
\small\begin{eqnarray}
\vartheta_{Inc}(x,t,k)&=&- E  \, t + k \,(x - x_{\0}),  \nonumber\\
\vartheta_{\R}(x,t,k) &=&- E  \, t - k \,(x + x_{\0}) + \Theta(k),\nonumber\\
\vartheta_{\A}(x,t,k) &=&- E  \, t - k \,x_{\0} + q  (x  - L) + \Theta(k),\nonumber\\
\vartheta_{\B}(x,t,k) &=&- E  \, t - k \,x_{\0}- q  (x - L) + \Theta(k),\nonumber\\
\vartheta_{\T}(x,t,k) &=&- E  \, t + k (x - x_{\0} - L)  + \Theta(k).
\label{p7}
\end{eqnarray}\normalsize
Meanwhile, these time-dependencies must be {\em carefully interpreted}.
In the Fig.(\ref{fig1A}), we illustrate the wave packet diffusion analytically described by the results of Eq.~(\ref{p6C}).
The ``pictures'' display the wave function in the {\em proximity} of the barrier for suitably chosen times.
By taking separately the right ($\alpha$) and left ($\beta$) moving components in region II, independent of the value of $\Theta(k)$, we can observe that
\small\begin{equation}
\Delta t_{\A} =  t_{\A}(L) -  t_{\A}(0) = \Delta t_{\B} =  t_{\B}(0) -  t_{\B}(L)= \left.\frac{L}{v_q} \right|_{k = k_{\0}}
\label{p13}
\end{equation}\normalsize
which correspond to transit time values ``classically'' expected.

However, by considering the information carried by the wave packet peaks, we would have a complete time discontinuity at $x = 0$ represented by Eq.~(\ref{p10A}) and by the fact that
\small\begin{equation}
t_{\A}(0) = - \left.\left[\frac{x_{\0}}{v_k} + \frac{L}{v_{q}}
 - \frac{\Theta^{\prime}(k)}{v_k} \right] \right|_{k = k_{\0}} ~~~\neq~~~t_{Inc}(0),
\label{p15}
\end{equation}\normalsize
i. e. the $\alpha$ wave could appear in region II at a time $t_{\A}(0)$ before (or after) the incident wave arrives at $x = 0$.
The results expressed by Eqs.~(\ref{p10A}) and (\ref{p15}) which are illustrated in Fig.~\ref{fig1A} immediately imply the non-conservation of probabilities since they lead to discontinuity points ($x=0$ and/or $x=L$) for the wave functions, which are Schr\"odinger equation solutions, , and for its spatial derivative.
The normalized squared modulus of this wave functions represents the spatial probability distribution of finding the corresponding propagating particle.
At $x=0$ (and/or $x=L$) the incoming flux of probabilities is different from the outgoing flux, which, from basic definitions of quantum mechanics continuity equation \cite{Coh77}, leads to the non-conservation of probabilities.
It is, by principle, unacceptable.

To be more accurate with such analysis and clear up some dubious points, it should be more convenient to verify two simple particular cases.
In order to simplify the calculations, we set $x_{\0} = 0 $ and we choose $q_{\0} \,  L = n  \pi ~~ (n = 1,2,3,...)$ so that we can write,
\small\begin{equation}
\Theta(k_{\0}) = 0,
~~~~ R(k_{\0}) = 0, ~~~~T(k_{\0}) = 1
\label{p16}
\end{equation}\normalsize
and
\small\begin{equation}
\Theta^{\prime}(k) = \frac{m \, L}{q_{\0}}
\frac{k^{\2}_{\0} + q^{\2}_{\0} }{2 k_{\0} q_{\0} } > \frac{m\,  L}{q_{\0} }
\label{p19A}
\end{equation}\normalsize
so that
\small\begin{equation}
t_{\A}(0) = \frac{m\,  L}{q_{\0} }
\frac{\left(k_{\0} - q_{\0} \right)^{\2}}{2 k_{\0} q_{\0} } > 0.
\end{equation}\normalsize
Otherwise, if we choose $q_{\0} \,  L = (n + \frac{1}{2})  \pi  $,
we shall obtain,
\small\begin{eqnarray}
&&\Theta(k) = \frac{\pi}{2},~~~~
R(k_{\0}) = \frac{k^{\2}_{\0} - q^{\2}_{\0} }{k^{\2}_{\0} + q^{\2}_{\0} },
~~~~T(k_{\0}) = \frac{2 k_{\0} \, q_{\0} }{k^{\2}_{\0} + q^{\2}_{\0} }
\exp{\left[i\left(\frac{\pi}{2} - k_{\0} \, L\right)\right]}
\label{p19C}
\end{eqnarray}\normalsize
and
\small\begin{equation}
\Theta^{\prime}(k) = \frac{m \, L}{q_{\0} }\frac{2 k_{\0} \,  q_{\0} }
{k^{\2}_{\0} + q^{\2}_{\0} } < \frac{m \, L}{q_{\0} }
\label{p19D}
\end{equation}\normalsize
which gives
\small\begin{equation}
t_{\A}(0) = -\frac{m L}{q_{\0} }
\frac{\left(k_{\0} - q_{\0} \right)^{\2}}{k^{\2}_{\0} + q^{\2}_{\0} } < 0.
\end{equation}\normalsize
The latter {\em negative} value illustratively corroborates with the absurd possibility of the appearance of peak associated with the $\alpha$ wave occurring before the arrival of the peak of the incident wave packet at $x = 0$, a clear representation of discontinuity at $x = 0$.
As we have stated before, the phase derivative $\Theta^{\prime}(k)$ illustrated in Fig.(\ref{fig0A}) does not present the adequate behavior for applying the SPM accurately.
In spite of this, it is currently ignored in the literature.

\subsection{Analytic solution for multiple peak decomposition}

Differently from the analytical analysis presented above, the numerical simulations of a wave packet diffusion by the potential barrier shows the appearance of multiple peaks due to the two reflective interfaces at $x = 0$ and $x = L$ (see the figures).
In fact, numerical calculations automatically preserve probabilities, at least to within the numerical errors.
This observation suggest a new analysis and subsequent interpretation/quantification of the ambiguities presented in the previous section.
In order to correctly apply the SPM and accurately obtain the position of the peak of the propagating wave packet with an accurate time dependence, we are constrict to solve the continuity constraints of the Schr\"odinger equation at each potential discontinuity point $x = 0$ and $x = L$ by considering multiple successive reflections and transmissions.
The phenomenon was already described in \cite{Ber04} and consists in an incoming wave of unitary amplitude with momentum distribution centered at $k_{\0}$ which reaches the interface $x = 0$ where, at an instant $t = t_{\0}$, is decomposed into a reflected wave component of amplitude (see the diagram) $R_{\1}$ and a transmitted wave component of amplitude $\alpha_{\1}$.
The transmitted wave continues propagating until it reaches $x = L$ where, at $t = t_{\1}$, it is now decomposed into a reflected wave component of amplitude $\beta_{\1}$ and a transmitted wave component of amplitude $T_{\1}$.
That new reflected wave component comes back to $x = 0$ where, at $t = t_{\2}$, it is again decomposed into another reflected wave component of amplitude $\alpha_{\2}$ and transmitted wave component of amplitude $R_{\2}$ which will be added to $R_{\1}$.
This iterative process continues for infinity times as we can see in the following diagram,
\footnotesize
\[
\begin{array}{l}
\begin{array}{r|l c r|l }
\exp[i\,k\,x] +  R_{\1}  \exp[-i\,k\,x] & \, \alpha_{\1}  \exp[i\,q\,x]&
 & \alpha{\1} \exp[i\,q\,x] \,+\,  \beta{\1}  \exp[-i\,q\,x]  & \, T_{\1} \exp[i\,k\,x]\\
 R_{\2}  \exp[-i\,k\,x] & \, \alpha{\2}  \exp[i\,q\,x] \,+\,  \beta{\1}  \exp[-i\,q\,x]&
 & \hspace*{0.5cm} \alpha{\2} \exp[i\,q\,x] \,+ \, \beta{\2}  \exp[-i\,q\,x]
& \, T_{\2} \exp[i\,k\,x]\\
\vdots \hspace*{1.2cm} &\hspace*{.75cm} \vdots \hspace*{.95cm}+ \hspace*{.7cm}
\vdots \hspace*{1.2cm}&  & \vdots \hspace*{.9cm}+ \hspace*{.7cm}
\vdots \hspace*{1.2cm}& \hspace*{.6cm}\vdots
\\
R_{\n}  \exp[-i\,k\,x] & \, \alpha_{\n} \exp[i\,q\,x] +  \beta_{\n - \1} \exp[-i\,q\,x]&
 & \hspace*{.5cm} \alpha_{\n} \exp[i\,q\,x] +  \beta_{\n}  \exp[-i\,q\,x]
& \, T_{\n} \exp[i\,k\,x]
\end{array}\\
\hspace*{3.35cm} x = 0 \hspace*{8.4cm} x = L
\end{array}
\]
\normalsize
The continuity constraints over $\psi(x,t)$ for each potential step at $x = 0$ and $x = L$ determine the coefficients
\small\begin{eqnarray}
&& R_{\1} = \frac{k - q}{k + q}\, ,~~~~ \alpha{\1}=
\frac{2\,k}{k + q}\, ,~~~~ \, \, \, \, \beta{\1}= \frac{2\,k (q -
k)}{(k + q)^{\2}}\, \exp[2 \,i\,q\,L\,]\, ,\hspace{5cm}\nonumber\\
&&T_{\1}=
\frac{4 \,k  q}{(k + q)^{\2}}\,\exp[i\,(q - k)\,L] \, ,~~~~
R_{\2} =\frac{q}{k}\, \alpha{\1} \, \beta{\1},
\label{p19BB}
\end{eqnarray}\normalsize
and establish the recurrence relations
\small\begin{eqnarray}
\frac{R_{\n + \2}}{R_{\n + \1}} = \frac{\alpha_{\n + \1}}{\alpha_{\n}} =
\frac{\beta_{\n + \1}}{\beta_{\n}} = \frac{T_{\n + \1}}{T_{\n}} =\left(
\frac{k - q}{k + q}\right)^{\2} \, \exp [2\, i\,q\,L\,] \hspace*{1cm}
\mbox{\small $n=1,2,\dots$}\, \,,\hspace*{2.4cm}
\label{p19B}
\end{eqnarray}\normalsize
which allow us to write the sum of the coefficients $R_{\n}$, $\alpha_{\n}$,$\beta_{\n}$, and $T_{\n}$ as
\small\begin{eqnarray}
R      &=&  \sum_{n = \1}^{\infty}R_{\n} =
R_{\1} + R_{\2}  \left[ 1 - \left(
\frac{k - q}{k + q}\right)^{\2} \, \exp [2\, i\,q\,L\,] \right]^{- \1}\,
,\nonumber \\
 \alpha
&=& \sum_{n = \1}^{\infty}\alpha{\n} = \alpha{\1}\left[ 1 - \left( \frac{k
- q}{k + q}\right)^{\2} \, \exp [2\, i\,q\,L\,] \right]^{- \1}\, \, ,\nonumber\\
 \beta
&=& \sum_{n = \1}^{\infty}\beta{\n} = \beta{\1}\left[ 1 - \left( \frac{k
- q}{k + q}\right)^{\2} \, \exp [2\, i\,q\,L\,] \right]^{- \1} \, \, ,\nonumber\\
 T
&=& \sum_{n = \1}^{\infty}T_{\n}\, = \, T_{\1} \left[ 1 - \left(
\frac{k - q}{k + q}\right)^{\2} \, \exp [2\, i\,q\,L\,] \right]^{-
\1}\, \, .
\label{p19}
\end{eqnarray}\normalsize
The above summations reproduce exactly the expressions in Eq.~(\ref{p5}).
In this form the interpretation is easy.
$R_{\1}$ represents the first reflected wave (it has no time delay since it is real).
$R_{\2}$ represents the second reflected wave and, as a consequence of the continuity condition at $x = 0$, it is the sum, in region II, of the first left-going wave ($\beta{\1}$) and the second right-going amplitude $(\alpha{\2})$, i.e.
\[ R_{\2}= \alpha{\2} + \beta{\1} \equiv \frac{q}{k}\, \alpha{\1} \,
\beta{\1}\, \, .\]
This structure is that given by considering two ``step functions'' back-to-back.
Thus at each interface the ``reflected'' and ``transmitted'' waves are instantaneous i. e. without any delay time.

Now we can calculate the time at which each transmitted and/or reflected wave appear by applying the SPM for each component of the total transmitted $T$ or reflected $R$ coefficients.
It will give us the recurrence relation
\small\begin{equation}
t_{n} = t_{n-1} + \frac{m (x -L)}{q_{\0}}
\label{2p29}
\end{equation}\normalsize
which coincides with the ``classically'' predicted value for the velocity of the particle above the barrier.
Indeed the SPM {\em applied separately} to each term in the above series expansion for $R$ yields delay multiples of $2\left.(\mbox{d}q/\mbox{d}E)\right|_{q = q_{\0}}\,L= 2 (m /q_{\0})\,L$.
This agrees perfectly with the fact that since the peak momentum in region II is $q_{\0}$, the $\alpha$ and $\beta$ waves have group velocities of $q_{\0}/m$ and hence transit times (one way) of $(m / q_{\0})\,L$.
The first transmitted peak appears (according to this version of the SPM) after a time $(m / q_{\0})\, L$, in perfect agreement with the above interpretation.

Is this compatible with probability conservation?
It is because of the following identity
\small\begin{equation}
\sum_{n = \1}^{\infty} \left( |R_{\n}|^{\2} + |T_{\n}|^{\2} \right)
=1\, \, .
\end{equation}\normalsize
This result is by no means obvious since it coexists with the well-known result, from the plane wave
analysis,
\small\begin{equation}
|R|^{\2} + |T|^{\2} = |\sum_{n = \1}^{\infty}R_{\n} \,|^{\2} + |
\sum_{n = \1}^{\infty}T_{\n} \,|^{\2}= 1\, \, .
\end{equation}\normalsize
To conclude, we can state that the conditions for applying the SPM in scattering problems of such a type depend upon the correct manipulation of the amplitude.
A posteriori this seems obvious but the point is that unless we know the number of separate peaks the SPM is inaccurate.
There is, however, a converse to this question.
If the modulating function is such that two or more wave packets overlap, then we cannot treat them separately without considering all the interference effects.
The above-barrier analysis is simply a particular example of this ambiguity.

\subsection{The analytical formula}

\begin{figure}
\vspace{-0.8cm}
\begin{center}
\epsfig{file= 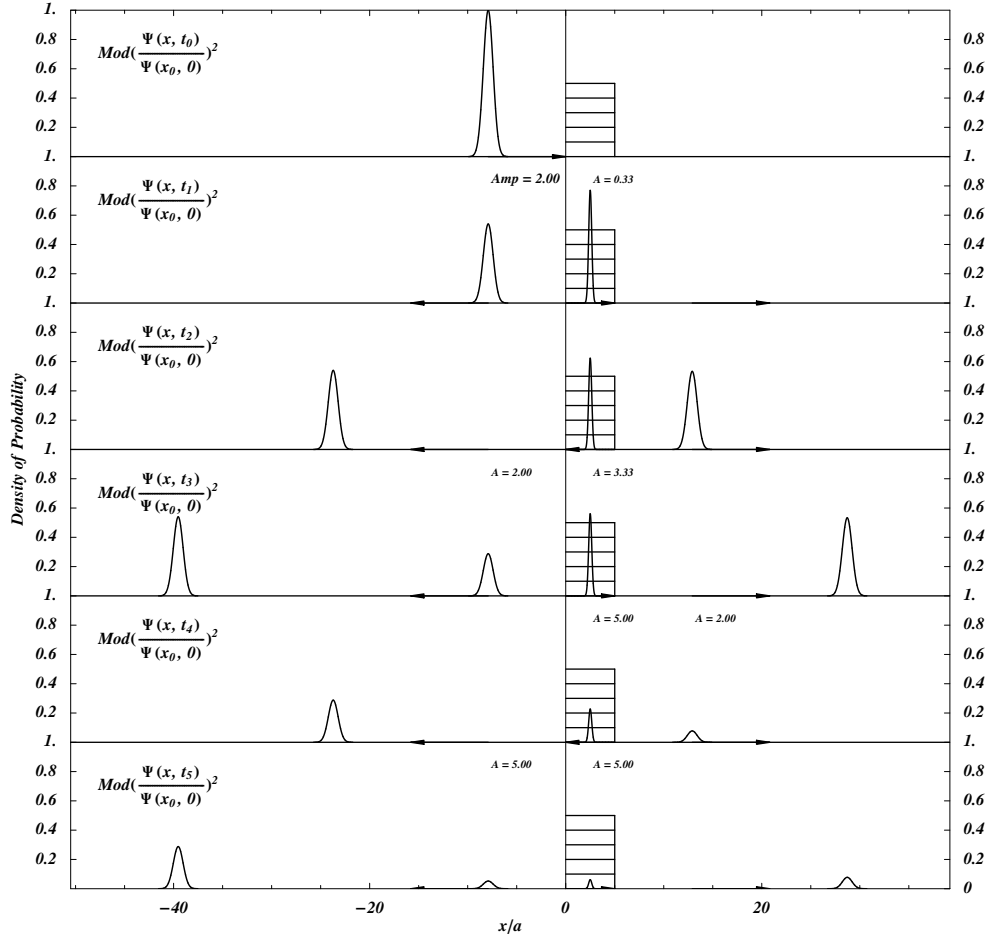, width = 13 cm}
\end{center}
\vspace{-0.8cm}
\caption{\footnotesize Multiple peak decomposition of a propagating wave packet in the above potential barrier scattering problem.
We have plotted the first few reflected and transmitted waves for times $t_{n} = (m a^{\2})[n\,(L/a)]/(a\, q_{\0})$ (with $n = 0,\,1,\,...,\,5$ and $m a^{\2} = 1$) in correspondence with Eq.~(\ref{2p29}), by assuming that the incoming wave packet starts at $x =- (k_{\0}\,L)/(2q_{\0})$.
The density of probabilities represents the collision of a wave packet of average width $a$ with a potential barrier $V_{\0}$ of width $L = 5\,a$.
Just for illustrating reasons, we have adopted $k_{\0} = (\sqrt{10} \,w)/3$ ($w a = 10000$) and we have printed the wave packet amplification multiplying factor ({\em A}) (individually adopted for visual convenience for each wave packet) when necessary.}
\label{fig3}
\end{figure}

Using the linear approximation (\ref{p6B}) for the momentum $q$ and the {\em Gaussian} momentum distribution (\ref{203}), and considering the expressions in (\ref{p19}) resulting from the multiple peak decomposition in (\ref{p6A}), newly with the same pertinent approximations used for obtaining (\ref{p6C}), we can analytically construct the following simplified expressions
\footnotesize\begin{eqnarray}
\psi_{\I}(x,t)&=&\varphi[x - x_{\0},\, t\,],\nonumber\\
\psi_{\R}(x,t) & = & \mbox{$\frac{k_{\0} - q_{\0}}{k_{\0} +
q_{\0}}$}\,\varphi[-\,x-x_{\0},\, t\,]+\nonumber\\
&&
\mbox{$\frac{4k_{\0}q_{\0}(q_{\0}-k_{\0})}{(k_{\0} +
q_{\0})^{\3}}$} \, \exp\left[-i\, \mbox{$\frac{2 w^{\2}}{q_{\0}}$}
\, L\right]\,\sum_{n = \0}^{\infty} \left( \mbox{$\frac{k_{\0} -
q_{\0}}{k_{\0} + q_{\0}}$} \, \exp\left[-i\,
\mbox{$\frac{w^{\2}}{q_{\0}}$} \, L\right] \right)^{\2 n}
\varphi[- x - x_{\0}+2(n+1)\mbox{$\frac{k_{\0}}{q_{\0}}$}\,L,\, t\,],\nonumber\\
\psi_{\A}(x,t) & = & \mbox{$\frac{2k_{\0}}{k_{\0} +
q_{\0}}$}\,\exp\left[-i\, \mbox{$\frac{w^{\2}}{q_{\0}}$} \,
x\right]\,\sum_{n = \0}^{\infty} \left( \mbox{$\frac{k_{\0} -
q_{\0}}{k_{\0} + q_{\0}}$} \, \exp\left[-i\,
\mbox{$\frac{w^{\2}}{q_{\0}}$} \, L\right] \right)^{\2 n}
\varphi[(x+2nL)\,\mbox{$\frac{k_{\0}}{q_{\0}}$} - x_{\0},\, t\,],\nonumber\\
\psi_{\B}(x,t) & = & \mbox{$\frac{2k_{\0}(q_{\0}-k_{\0})}{(k_{\0}
+ q_{\0})^{\2}}$}\,\exp\left[i\, \mbox{$\frac{w^{\2}}{q_{\0}}$}
\, (x-2L)\right]\,\sum_{n = \0}^{\infty} \left(
\mbox{$\frac{k_{\0} - q_{\0}}{k_{\0} + q_{\0}}$} \, \exp\left[-i\,
\mbox{$\frac{w^{\2}}{q_{\0}}$} \, L\right] \right)^{\2 n}
\varphi[(2nL + 2L - x)\,\mbox{$\frac{k_{\0}}{q_{\0}}$} - x_{\0},\, t\,],\nonumber\\
\psi_{\T}(x,t) & = & \mbox{$\frac{4k_{\0}q_{\0}}{(k_{\0} +
q_{\0})^{\2}}$}\, \exp\left[-i\, \mbox{$\frac{w^{\2}}{q_{\0}}$}
\, L\right]\,\sum_{n = \0}^{\infty} \left( \mbox{$\frac{k_{\0} -
q_{\0}}{k_{\0} + q_{\0}}$} \, \exp\left[-i\,
\mbox{$\frac{w^{\2}}{q_{\0}}$} \, L\right] \right)^{\2 n}
\varphi[x - x_{\0}-L+(2n+1)\mbox{$\frac{k_{\0}}{q_{\0}}$}\,L,\, t\,].\nonumber\\
\label{p25}
\end{eqnarray}\normalsize
which allow us to visualize the multiple peak decomposition illustrated in Fig.~\ref{fig3}.

\begin{figure}
\vspace{-0.8cm}
\begin{center}
\epsfig{file= 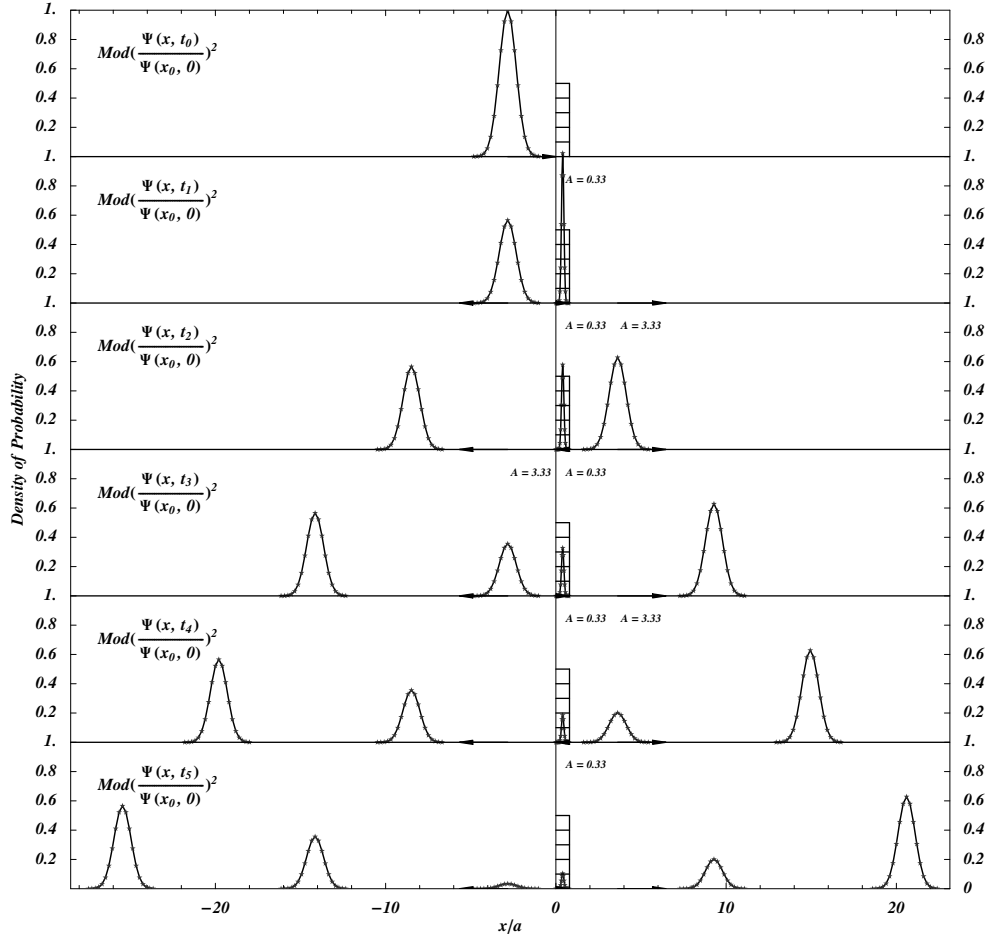, width = 13 cm}
\end{center}
\vspace{-0.8cm}
\caption{\footnotesize
The illustrative confront between the analytical (line) and the numerical (star symbols) results for reflected and transmitted waves when the condition $1 < k_{\0}/q_{\0} < \pi$ is satisfied and, consequently, both the analytical and the numerical results have to be equivalent.
The stars are from the numerical convolution of the plane-wave solution and the curves are from the analytical expressions for the first three $R_{n}$ and $T_{n}$.
The density of probabilities represents the collision of a wave packet of average width $a$ with a potential barrier $V_{\0}$ of width $L = 0.8\, a$.
Again for illustrating reasons, we have adopted $k_{\0} = (5\sqrt{2}\,w)/7$ and we have printed the wave packet amplification multiplying factor ({\em A}) when necessary.
The reason for choosing a so large value for the parameter $w a$ ($w a = 10000$) in all these figures is to guarantee the convergence between the analytical and the numerical results for all the decomposed peaks.}
\label{fig4}
\end{figure}

Comparing with the exact numerical results, the analytic approximations expressed in (\ref{p25}) will be valid only under certain restrictive conditions.
To explain such a statement, when we analytically solve the integrals like (\ref{p6A}), the exponential $\exp{[2\, i \, q \, L]}$ which appears in the recurrence relations (\ref{p19B}) are approximated by
\small\begin{equation}
\exp{[2\, i \, q_{\0} \, L + 2\, i \, (k - k_{\0})\frac{k_{\0}}{q_{\0}}]}.
\end{equation}\normalsize
In this case, the above analytical expressions can be obtained only when the exponential function does not oscillate in the interval of relevance of the envelop {\em Gaussian} function $g(k - k_{\0})$, i. e. when $\Delta q \, L < \pi$, which can be expressed in terms of the wave packet width as
\small\begin{equation}
\Delta q \, L \, \approx\,  \left.\frac{dq}{dk}\right|_{k = k_{\0}} \, \Delta k \, L \,  \approx\,  \frac{k_{\0}}{q_{\0}} \frac{L}{a} < \pi.
\label{p26}
\end{equation}\normalsize
where we have used $\Delta k \, \approx\, 1/a$.
The above constraint implicitly carries a very peculiar character: the multiple peak decomposition can also be evidenced when the wave packet width $a$ is larger than the potential barrier width $L$, i. e. when we assume the relation $k_{\0}/q_{\0}$ is in the interval $1 < k_{\0}/q_{\0} < \pi$ we can have propagating wave packets with $a > L$ satisfying the requirements for the multiple peak resolution (\ref{p25}).
To illustrate this possibility, in Fig.~\ref{fig4} we plot the density of probabilities representing the collision of a wave packet of average width $a$ with a potential barrier $V_{\0}$ of width $L = 0.8\, a$. We compare the analytical (line) with the numerical (star symbols) results from which we can evidently observe the accurate equivalence between them since we are respecting the condition (\ref{p26}).
Both the figures (\ref{fig3}) and (\ref{fig4}) display the square of the wave function in the {\em proximity} of the barrier for suitably chosen times.
One clearly sees the appearance of multiple peaks due to the two reflective interfaces at $x=0$ and $x=L$.

\section{Limitations on the tunneling analysis}

Generically speaking, the SPM can be successfully utilized for describing the motion of the center of a wave packet constructed in terms of a momentum distribution $g(k - k_{\0})$ which has a pronounced peak around $k_{\0}$.
In case of a potential barrier given by,
\small\begin{equation}
V(x) = \left\{\begin{array}{cll} V_o && ~~~~x \in \mbox{$\left[- L/2, \, L/2\right]$}\\ &&\\ 0&& ~~~~x \in\hspace{-0.3cm}\slash\hspace{0.1cm}\mbox{$\left[- L/2, \, L/2\right]$}\end{array}\right.
\label{2p60}
\end{equation}\normalsize
it is well known that the transmitted wave packet solution ($x \geq L/2 $) calculated by means of the Schr\"odinger formalism is given by \cite{Coh77}
\small\begin{equation}
\psi^{\T}(x,t) = \int_{_{\0}}^{^{w}}\frac{dk}{2\pi} \, g(k - k_{\0}) \, |T(k, L)|\,
\exp{\left[ i \, k \,(x - L/2) - i \, \frac{k^2}{2\,m} \, t +
 i \,\Theta(k, L)\right]}
\end{equation}\normalsize
where, in case of tunneling, the transmitted amplitude is written as
\small\begin{equation}
|T(k, L)| =
\left\{1+ \frac{w^4}{4 \, k^2 \, \rho^{\2}(k)}
\sinh^2{\left[\rho(k)\, L \right]}\right\}^{-\frac{1}{2}},
\label{1}
\end{equation}\normalsize
and the phase shift is described in terms of
\small\begin{equation}
\Theta(k, L) = \arctan{\left\{\frac{2\, k^2 - w^2}
{k \, \rho(k)}
\tanh{\left[\rho(k) \, L \right]}\right\}},
\label{502}
\end{equation}\normalsize
for which we have made explicit the dependence on the barrier length $L$ and we have adopted $\rho(k) = \left(w^2 - k^2\right)^{\frac{1}{2}}$ with $w = \left(2\, m \,V_{\0}\right)^{\frac{1}{2}}$ and $\hbar = 1$.
Without thinking over an eventual distortion that $|T(k, L)|$ causes to the supposedly symmetric function $g(k - k_{\0})$, when the stationary phase condition is applied to the phase of Eq.~(\ref{1}) we obtain
\small\begin{eqnarray}
\frac{d}{dk}\left.\left\{k \,(x - L/2) - \frac{k^2}{2\,m} \, t
+ \Theta(k, L)\right\}\right|_{_{k = k_{\mbox{\tiny max}}}}
&=&0 ~~~~\Rightarrow\nonumber\\
  x - L/2 - \frac{k_{\mbox{\tiny max}}}{m} \, t +
\left.\frac{d\Theta(k, L)}{dk}\right|_{_{k = k_{\mbox{\tiny max}}}} &=& 0.
\label{3}
\end{eqnarray}\normalsize
The above result is frequently adopted for calculating the transit time $t_{T}$ of a transmitted wave packet when its peak emerges at $x = L/2$,
\small\begin{equation}
t_{T} =\frac{m}{k_{\mbox{\tiny max}}}\left.\frac{d\Theta(k, \alpha_{(\L)})}{dk}\right|_{_{k = k_{\mbox{\tiny max}}}} =
\frac{2\,m \, L}{k_{\mbox{\tiny max}} \,\alpha }
\left\{\frac{w^4\,\sinh{(\alpha)}\cosh{(\alpha)}
-\left(2\, k_{\mbox{\tiny max}}^2 - w^2 \right)k_{\mbox{\tiny max}}^2 \,\alpha }
{4\, k_{\mbox{\tiny max}}^2 \,\left(w^2 - k_{\mbox{\tiny max}}^2 \right)  +
w^4\,\sinh^2{(\alpha)}}\right\}
\label{4}
\end{equation}\normalsize
where we have defined the parameter $\alpha = \left(w^2 - k_{\mbox{\tiny max}}^2 \right)^{\frac{1}{2}}\, L$.
The concept of {\em opaque} limit is introduced when we assume that $k_{\mbox{\tiny max}}$ is independent of $L$ and then we make $\alpha$ tends to $\infty$ \cite{Jak98}.
In this case, the transit time can be rewritten as
\small\begin{equation}
t^{^{OL}}_T = \frac{2\,m}{k_{\mbox{\tiny max}}\,\rho(k_{\mbox{\tiny max}})}.
\label{5}
\end{equation}\normalsize
In the literature, the value of $k_{\mbox{\tiny max}}$ is frequently approximated by $k_{\0}$, the maximum of $g(k - k_{\0})$, which, in fact, does not depend on $L$ and could lead us to the transmission time superluminal interpretation \cite{Jak98,Olk92,Esp03} of (\ref{5}).
To clear up this point, we notice that when we take the {\em opaque} limit with $L$ going to $\infty$ and $w$ fixed as well as with $w$ going to $\infty$ and $L$ fixed, with $k_{\0} < w$ in both cases, the expression (\ref{5}) suggests times corresponding to a transmission process performed with velocities higher than $c$ \cite{Jak98}.
Such a superluminal interpretation was extended to the study of quantum tunneling through two successive barriers separated by a free region where it was theoretically verified that the total traversal time does not depend not only on the barrier widths, but also on the free region extension \cite{Olk92}.
Besides, in a subsequent analysis, the same technique was applied to a problem with multiple successive barrier where the tunneling process was designated as a highly non-local phenomenon \cite{Esp03}.

It would be perfectly acceptable to consider $k_{\mbox{\tiny max}} = k_{\0}$ for the application of the stationary phase condition when the momentum distribution $g(k - k_{\0})$ centered at $k_{\0}$ is not modified by any bound condition.
This is the case of the incident wave packet before colliding with the potential barrier.
Otherwise, the procedure for obtaining all the above results for the the transmitted wave packet does not take into account the bounds and enhancements imposed by the analytical form of the transmission coefficient.

Hartman himself had already noticed that the

\vspace{0.3cm}
{\it`` Discussion of the time dependent behavior of a wave packet is complicated by its diffuse or spreading nature; however, the position of the peak of a {\em symmetrical} packet can be described with some precision''}
\vspace{0.3cm}

The origin of the Hartman effect, and the above-illustrated saturation of group delay with barrier length, has been a mystery for decades.
The phenomenon exists in the tunneling of all kinds of waves, be they matter waves, electromagnetic waves, or sound
waves.
Experimentally it has been observed using electromagnetic waves and sound waves and there is no question as
to its existence.
If one interprets the group delay in tunneling as a transit time then the Hartman effect naturally leads to superluminal and unbounded group velocities.
However, to perform the correct analysis, we should calculate the right value of $k_{\mbox{\tiny max}}$ to be substituted in Eq.~(\ref{4}) before taking the {\em opaque} limit.
We have to consider the relevant amplitude for the transmitted wave as the product of a symmetric momentum distribution $g(k - k_{\0})$ which describes the {\em incoming} wave packet by the modulus of the transmission amplitude $T(k, L)$ which is a crescent function of $k$.
The maximum of this product representing the transmission modulating function would be given by the solution of the equation
\small\begin{eqnarray}
g(k - k_{\0}) \,\left|T(k, L)\right|\,\left[\frac{g^{\prime}(k - k_{\0})}{g(k - k_{\0})}+
\frac{\left|T(k, L)\right|^{\prime}}{\left|T(k, L)\right|}\right] &=& 0
\label{3p42B}
\end{eqnarray}\normalsize
Obviously, the peak of the modified momentum distribution is shifted to the right of $k_{\0}$ so that $k_{\mbox{\tiny max}}$ have to be found in the interval $]k_{\0}, w[$.
Moreover, we confirm that $k_{\mbox{\tiny max}}$ presents an implicit dependence on $L$ as we can demonstrate by the numerical results presented in {\it Table 1} where we have found the maximum of $g(k - k_{\0}) \, |T(k, L)|$ by assuming $g(k - k_{\0})$ is a {\em Gaussian} function almost completely comprised in the interval $[0, w]$ given by
\small\begin{equation}
g(k - k_{\0}) = \left(\frac{a^2}{2 \, \pi}\right)^{\frac{1}{4}}
\exp{\left[-\frac{a^2 (k -k_{\0})^2}{4}\right]}.
\label{6}
\end{equation}\normalsize
By increasing the value of $L$ with respect to $a$, the value of $k_{\mbox{\tiny max}}$ obtained from the numerical calculations to be substituted in Eq.~(\ref{4}) also increases until $L$ reaches certain values for which the modified momentum distribution becomes unavoidably distorted.
In this case, the relevant values of $k$ are concentrated around the upper bound value $w$.
We shall show in the following that the value of $L$ which starts to distort the momentum distribution can be analytically obtained in terms of $a$.

Now, if we take the {\em opaque} limit of $\alpha$ by fixing $L$ and increasing $w$, the above results immediately ruin the superluminal interpretation upon the result of Eq.~(\ref{4}) since $t^{^{OL}}_T$ tends to $\infty$ when $k_{\mbox{\tiny max}}$ is substituted by $w$.

Otherwise, when $w$ is fixed and $L$ tends to $\infty$, the parameter $\alpha$ calculated at $k = w$ becomes indeterminate.
The transit time $t_T$ still tends to $\infty$ but now it exhibits a peculiar dependence on $L$ which can be easily observed by defining the auxiliary function
\small\begin{equation}
G(\alpha) =
\frac{\sinh{(\alpha)}\cosh{(\alpha)} - \alpha}
{\sinh^2{(\alpha)}}
\label{11}
\end{equation}\normalsize
which allow us to write
\small\begin{equation}
t^{\alpha}_T = \frac{2\, m\, L}{w \, \alpha}\,
G(\alpha).
\label{12}
\end{equation}\normalsize
When $\alpha \gg 1$, the transmission time always assume infinite values with an asymptotic dependence on $\left(w^2 - k^2 \right)^{-\frac{1}{2}}$,
\small\begin{equation}
t^{\alpha}_T \approx  \frac{2\, m }{w \, \left(w^2 - k^2 \right)^{\frac{1}{2}}} \rightarrow \infty.
\label{13}
\end{equation}\normalsize
Only when $\alpha$ tends to $0$ we have an explicit linear dependence on $L$ given by
\small\begin{equation}
t^{0}_T = \frac{2\, m \, L}{w}\,\lim_{\alpha \rightarrow 0}
{\left\{\frac{G(\alpha)}{\alpha}\right\}}
= \frac{4\, m \, L}{3 \, w}
\label{14}
\end{equation}\normalsize

In addition to the above results, the transmitted wave must be carefully studied in terms of the rapport between the barrier extension $L$ and the wave packet width $a$.
For very thin barriers, i. e. when $L$ is much smaller than $a$, the modified transmitted wave packet presents substantially the same form of the incident one.
For thicker barriers, but yet with $L < a$, the peak of the {\em Gaussian} wave packet modulated by the transmission coefficient is shifted to higher energy values, i. e.  $k_{\mbox{\tiny max}} > k_{\0}$ increases with $L$.
For very thick barriers, i. e. when $L > a$, we are able to observe that the form of the transmitted wave packet is badly distorted with the greatest contribution coming from the Fourier components corresponding to the energy $w$ just above the top of the barrier in a kind of {\em filter effect}.
We observe that the quoted distortion starts to appear when the modulated momentum distribution presents a {\em local maximal} point at $k = w$ which occurs when
\small\begin{equation}
\left.\frac{d}{dk}\left[g(k - k_{\0}) \,
\left|T(k, L)\right|\right]\right|_{k = w} > 0
\label{a}.
\end{equation}\normalsize
Since the derivative of the {\em Gaussian} function $g(k - k_{\0})$ is negative at $k = w$, the Eq.~(\ref{a}) gives the relation
\small\begin{equation}
- \frac{g^{\prime}(w - k_{\0})}{g(w - k_{\0})} < \lim_{k \rightarrow w}
{\left[\frac{T^{\prime}(k, L)}{T(k, L)}\right]}
= \frac{w \, L^2}{4}\frac{\left(1 + \frac{w \, L^2}{3}\right)}
{\left(1 + \frac{w \, L^2}{4}\right)} < \frac{w \, L^2}{3}
\label{b}
\end{equation}\normalsize
which effectively represents the inequality
\small\begin{equation}
\frac{a^2}{2}\,(w - k_{\0}) < \frac{w \,L^2}{3}~~\Rightarrow~~
L > \sqrt{\frac{3}{2}}\,a\, \left(1 - \frac{k_{\0}}{w}\right).
\label{c}
\end{equation}\normalsize

\begin{figure}
\centerline{\psfig{file=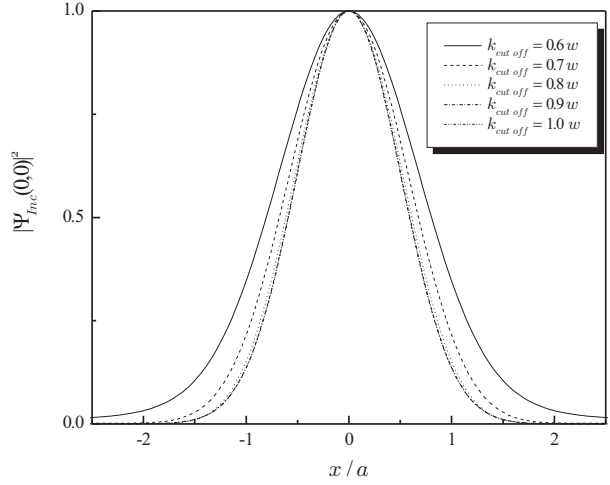, width=9 cm}}
\vspace*{8pt}
\caption{Dependence of the wave packet shape on the {\em cut off} value of a momentum distribution centered around $k_{o} = 0.5 w$ with the values of $k$ comprised between $0$ and $k_{\mbox{\tiny \em{cut off}}}$.
\label{fig2A}}
\end{figure}

Due to the {\em filter effect}, the amplitude of the transmitted wave is essentially composed by the plane wave components of the front tail of the {\em incoming} wave packet which reaches the first barrier interface before the peak arrival.
Meanwhile, only whether we had {\em cut} the momentum distribution {\em off} at a value of $k$ smaller than $w$, i. e. $k \approx (1 - \delta) w$, the superluminal interpretation of the transition time (\ref{5}) could be recovered.
In this case, independently of the way as $\alpha$ tends to $\infty$, the value assumed by the transit time would be approximated by $t^{\alpha}_{T} \approx 2 \,m / w \, \delta$ which is a finite quantity.
Such a finite value would confirm the hypothesis of {\em superluminality}.
However, the {\em cut off} at $k \approx (1 - \delta) w$ increases the amplitude of the tail of the incident wave as we can observe in Fig.~\ref{fig2A}.
It contributes so relevantly as the peak of the incident wave to the final composition of the transmitted wave and creates an ambiguity in the definition of the {\em arrival} time.

To summarize, at this point we are particularly convinced that the use of a step-discontinuity to analyze signal transmissions in tunneling processes deserves a more careful analysis than the immediate application of the stationary phase method since we cannot find an analytic-continuation between the {\em above} barrier case solutions and the {\em below} barrier case solutions.
A suggestive possibility can ask for the use of the multiple peak decomposition technique developed for the above-barrier diffusion problem \cite{Ber04} presented in section III.
Thus, in the same framework, we suggest in the following section a suitable way for comprehending the conservation of probabilities for a very particular tunneling configuration where the asymmetry presented here is eliminated, and the phase times can be accurately calculated.

\section{The symmetrical tunneling configuration}

Due to the factual influence of the amplitude of the transmitted wave in determining the right position of a tunneling particle, we can introduce an alternative analysis where the possibility of multiple peak decomposition \cite{Ber04} may be taken into account.
By means of such an experimentally verifiable exercise, we shall be able to understand how the {\em filter effect} can analytically affect the calculations of transit times in the tunneling process.

In order to recover the scattered momentum distribution symmetry conditions for accurately applying the SPM, we assume a symmetrical colliding configuration of two wave packets traveling in opposite directions.
By considering the same barrier represented in (\ref{2p60}), we solve the Schr\"odinger equation for a plane wave component of momentum $k$ for two identical wave packets symmetrically separated from the origin $x = 0$, at time of collision $t = - (m L) /(2 k_{\0})$ chosen for mathematical convenience, where we assume they perform a totally symmetric simultaneous collision with the potential barrier.
The wave packet reaching the left(right)-side of the barrier is represented by
\small\begin{equation}
\psi^{\L(\R)}(x,t) = \int_{_{\mi\infty}}^{^{\pl\infty}}dk \, g(k - k_{\0})\phi^{\L(\R)}(k,x)\, \exp{[- i \, E\,t]}
\end{equation}\normalsize
where we assume the integral can be naturally extended from the interval $[0,w]$ to the interval $[-\infty,\pl\infty]$ as a first approximation.
In the framework of the multiple peak decomposition \cite{Ber04}, we have suggested a suitable way for comprehending the conservation of probabilities where the asymmetric aspects previously discussed \cite{Ber06} could be totally eliminated.
By considering the same rectangular barrier $V(x)$, we solve the Schr\"odinger equation for a plane wave component of momentum $k$ for two identical wave packets symmetrically separated from the origin $x = 0$.
By assuming that $\phi^{\L(\R)}(k,x)$ are Schr\"odinger equation stationary wave solutions, when the wave packet peaks simultaneously reach the barrier (at the mathematically convenient time $t = - (m L) /(2 k_{\0})$) we can write
\small\begin{equation}
\phi^{\L(\R)}(k,x)=
\left\{\begin{array}{l l l l}
\phi^{\L(\R)}_{\1}(k,x) &=&
\exp{\left[ \pm i \,k \,x\right]} + R^{\L(\R)}(k,L)\exp{\left[ \mp i \,k \,x\right]}&~~~~x < - L/2\, (x > L/2),\nonumber\\
\phi^{\L(\R)}_{\2}(k,x) &=& \gamma^{\L(\R)}(k)\exp{\left[ \mp\rho  \,x\right]} + \beta^{\L(\R)}(k)\exp{\left[ \pm\rho  \,x\right]}&~~~~- L/2 < x < L/2,\nonumber\\
\phi^{\L(\R)}_{\3}(k,x) &=& T^{\L(\R)}(k,L)\exp{ \left[\pm i \,k \,x\right]}&~~~~x > L/2 \, (x < - L/2) .
\end{array}\right.
\label{510}
\end{equation}\normalsize
where the upper(lower) sign is related to the index $L$($R$) corresponding to the incidence on the left(right)-hand side of the barrier.
By assuming the conditions for the continuity of $\phi^{\L,\R}$ and their derivatives at $x = - L/2$ and $x = L/2$, after some mathematical manipulations, we can easily obtain
\small\begin{equation}
R^{\L,\R}(k,L) = \exp{\left[ - i \,k \,L \right]} \left\{\frac{\exp{\left[ i \, \theta(k)\right]} \left[1 - \exp{\left[ 2\,\rho(k) \,L\right]}\right]}{1 - \exp{\left[ 2\,\rho(k) \,L\right]}\exp{\left[ i\, 2\,\theta(k)\right]}}\right\}
\label{511}
\end{equation}\normalsize
and
\small\begin{equation}
T^{\L,\R}(k,L) = \exp{\left[ - i \,k \,L \right]} \left\{\frac{\exp{\left[\rho(k) \,L\right]}\left[1- \exp{\left[ 2\, i\, \theta(k)\right]}\right]}{1 - \exp{\left[ 2\,\rho(k) \,L\right]}\exp{\left[ i\, 2\,\theta(k)\right]}}\right\},
\label{512}
\end{equation}\normalsize
where
\small\begin{equation}
\theta(k) = \frac{ 2\, k \, \rho(k)}{2k^{\2} - w^{\2}}
\label{512B}
\end{equation}\normalsize
and $R^{\L(\R)}(k,L)$ and $T^{\R(\L)}(k,L)$ are intersecting each other.

\subsection{Symmetrized and antisymmetrized quantum tunneling configuration}

Since the above introduced colliding configuration is spatially symmetric, the symmetry operation corresponding to the $1 \leftrightarrow 2$ particle exchange can be parameterized by the position coordinate transformation $x \rightarrow -x$.
At the same time, it is easy to observe that
\small\begin{equation}
\phi^{\L(\R)}(k,x) = \phi^{\L(\R)}_{\1 \pl \2 \pl \3}(k,x) = \phi^{\R(\L)}_{\1 \pl \2 \pl \3}(k,-x) = \phi^{\R(\L)}(k,-x)
\label{512C}
\end{equation}\normalsize
where the $L \leftrightarrow R$ interchange is explicit.
Consequently, in case of analyzing the collision of two identical bosons, we have to consider a symmetrized superposition of the $L$ and $R$ wave functions,
\small\begin{equation}
\phi_{\pl}(k,x) = \phi^{\L}(k,x) + \phi^{\R}(k,x) = \phi^{\R}(k,-x) + \phi^{\L}(k,-x) = \phi_{\pl}(k,-x).
\label{512D}
\end{equation}\normalsize
Analogously, in case of analyzing the collision of two identical fermions (just taking into account the spatial part of the wave function), we have to consider an antisymmetrized superposition of the $L$ and $R$ wave functions,
\small\begin{equation}
\phi_{\mi}(k,x) = \phi^{\L}(k,x) - \phi^{\R}(k,x) = \phi^{\R}(k,-x) - \phi^{\L}(k,-x) = -\phi_{\mi}(k,-x).
\label{512E}
\end{equation}\normalsize
Thus the amplitude of the re-composed transmitted plus reflected waves would be given by $R^{\L,\R}(k,L) + T^{\R,\L}(k,L)$ for the symmetrized wave function $\phi_{\pl}$ and by $R^{\L,\R}(k,L) - T^{\R,\L}(k,L)$ for the antisymmetrized wave function $\phi_{\mi}$.
Reporting to the multiple peak decomposition \cite{Ber04} applied to such a pictorial symmetrical tunneling configuration \cite{Ber08C}, we can superpose the amplitudes of the intersecting probability distributions before taking their squared modulus in order to obtain
\small\begin{eqnarray}
R^{\L,\R}(k,L) \pm T^{\R,\L}(k,L) &=& \exp{\left[ - i \,k \,L \right]} \left\{\frac{\exp{\left[\rho(k) \,L\right]}\pm \exp{\left[ i\, \theta(k)\right]}}{1 \pm \exp{\left[ \rho(k) \,L\right]}\exp{\left[ i\, \theta(k)\right]}}\right\}
\nonumber\\
&=&
 \exp{\left\{ - i [k \,L - \varphi_{\pm}(k,L)]\right\}}
\label{513}
\end{eqnarray}\normalsize
with
\small\begin{equation}
\varphi_{\pm}(k,L) = - \arctan{\left\{\frac{2\,k\,\rho(k) \, \sinh{[\rho(k)\,L]}}{\left(k^{\2}-\rho^{\2}(k)\right)\cosh{[\rho(k)\,L]} \pm w^{\2}}\right\}}.
\label{514}
\end{equation}\normalsize\normalsize
where the {\em plus} sign is related to the results obtained for the a symmetrized superposition and the {\em minus} sign is related to the symmetrized superposition.
Independently of the odd or even symmetrization of the wave function, we observe from Eq.~(\ref{513}) that, differently from the previous standard tunneling analysis, by adding the intersecting amplitudes at each side of the barrier, we keep the original momentum distribution undistorted since $|R^{\L,\R}(k,L)\pm T^{\R,\L}(k,L)|$ is equal to one.
The previously pointed out incongruities which cause the distortion of the momentum distribution $g(k - k)$ are completely eliminated and we recover the fundamental condition for the applicability of the SPM for accurately determining the position of the peak of the reconstructed wave packet composed by reflected and transmitted superposing components.
The phase time interpretation can be, in this case, correctly quantified in terms of the analysis of the novel phase $\varphi_{\pm}(k, L)$ since the novel scattering amplitudes $g(k - k_{\0})|R^{\L,\R}\pm T^{\R,\L}| \approx g(k - k_{\0})$ maintains its previous symmetrical character.
The transmitted and reflected interfering amplitudes results in a unimodular function which just modifies the {\em envelop} function $g(k - k)$ by an additional phase. and the scattering phase time results in
\small\begin{equation}
t^{(\alpha)}_{T, \varphi_{\ppm}} =\frac{m }{k}\frac{d\varphi(k, \alpha)}{dk} =
\frac{2\,m\, L}{k\,\alpha}
\frac{w^{\2}\sinh{(\alpha)} \pm \alpha\,k^{\2}}{2\,k^{\2} - w^{\2} \pm w^{\2}\cosh{(\alpha)}}
\label{515}
\end{equation}\normalsize
where $k \rightarrow k_{\0}$, with $\alpha$ previously defined.
The {\em old} phase $\Theta(k, L)$ (Eq.~\ref{502}) appears when we treat separately the momentum amplitudes $T(k, L)$ and $R(k, L)$, which destroys the symmetry of the initial momentum distribution $g(k - k_{\0})$ by the presence of the multiplicative term $T(k, L)$ or $R(k, L)$, and the novel phase $\varphi(k, L)$ appears only when we sum the tunnneling/scattering amplitudes so that the symmetrical character of the initial momentum distribution is recovered (due to the result of Eq.~(\ref{513}).

We should assume that in the region inside the potential barrier, the reflecting and transmitting amplitudes should be summed before we compute the phase changes.
Obviously, it would result in the same phase time expression as represented by (\ref{515}).
In this case, the assumption of there (not) existing interference between the momentum amplitudes of the reflected and transmitted waves at the discontinuity points $x = -L/2$ and $x = L/2$ is purely arbitrary.
Consequently, it is important to reinforce the argument that such a possibility of interference leading to different phase time results is strictly related to the idea of using (or not) the multiple peak (de)composition in the region where the potential barrier is localized.
To illustrate the difference between the {\em standard} tunneling phase time $t^{(\alpha)}_{T}$ and the {\em symmetrical} scattering phase time $t^{(\alpha)}_{T, \varphi}$ we introduce the parameter $n = k^{\2}/w^{\2}$ and we define the {\em classical} traversal time $\tau_{\k} = (m L) /k$.
In this case, we can obtain the normalized phase times
\small\begin{equation}
t^{(\alpha)}_{T}
=
\frac{2  \tau_{\k}}{\alpha}\left\{
\frac{\cosh{(\alpha)}\sinh{(\alpha)} - \alpha\,n\left(2 n - 1\right)}{\left[4 n \left(1 - n\right)+\sinh^{\2}{(\alpha)}\right]}
\right\}
\label{517A}
\end{equation}\normalsize
and
\small\begin{equation}
t^{(\alpha)}_{T, \varphi_{\pm}}
=
\frac{2  \tau_{\k}}{\alpha}\left\{\frac{n\, \alpha \pm \sinh{(\alpha)}}{2n - 1 \pm\cosh{(\alpha)}}
\right\}.\label{517}
\end{equation}\normalsize

\begin{figure}
\vspace{-0.6 cm}
\centerline{\psfig{file=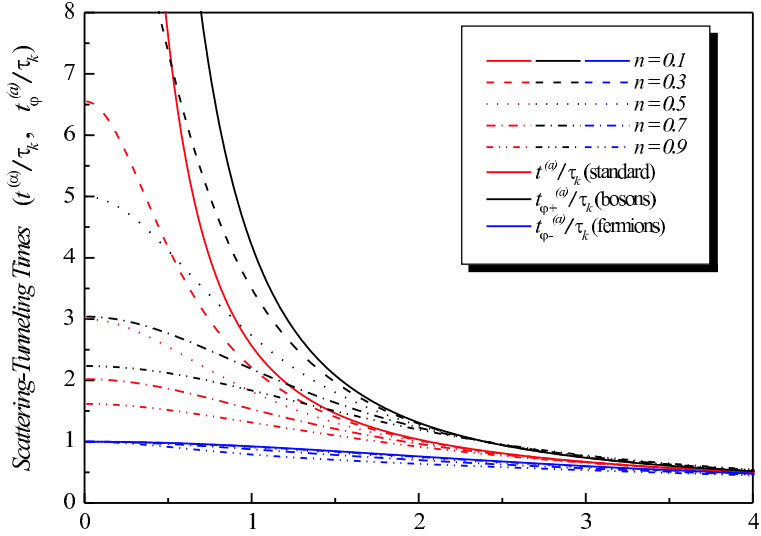,width=10cm}}
\vspace{-1 cm}
\caption{Normalized phase times for a {\em symmetrical} tunneling configuration of the symmetrized wave function representing the collision of two identical bosons (black lines) and the antisymmetrized wave function representing the collision of two identical fermions (red lines).
The results are plotted in comparison with the {\em standard} one-way direction tunneling phase times (blue lines).
These times can be understood as transit times in the units of the {\em classical} traversal time $\tau_{\k} = (m L) /k$.
All the above phase time definitions present the same asymptotic behavior.
\label{fig3A}}
\vspace{-0.4 cm}
\end{figure}

In order to illustrate the difference between the standard {\em tunneling} phase time $t_{T}$ and the alternative {\em scattering} phase time $t^{^{\varphi}}_{T}$ we introduce the new parameter $n = k^{\2}_{\mbox{\tiny max}}/w^{\2}$ and the {\em classical} traversal time $\tau = (m L) /k_{\mbox{\tiny max}}$ in order to define the rates
\small\begin{equation}
R_{T}(\alpha) = \frac{t_{T}}{\tau}=
\frac{2}{\alpha}\left\{
\frac{\cosh{(\alpha)}\sinh{(\alpha)} - \alpha\,n\left(2 n - 1\right)}{\left[4 n \left(1 - n\right)+\sinh^{\2}{(\alpha)}\right]}
\right\}\label{516}
\end{equation}\normalsize
and
\small\begin{equation}
R^{^{\varphi}}_{T}(\alpha) = \frac{t^{^{\varphi}}_{T}}{\tau}=
\frac{2}{\alpha}\left\{\frac{n\, \alpha + \sinh{(\alpha)}}{2n - 1 +\cosh{(\alpha)}}
\right\}\label{516B}
\end{equation}\normalsize
which are plotted in the Fig.(\ref{fig3A}) for some discrete values of $n$ varying from $0.1$ to $0.9$, from which we can obtain the most common limits given by
\small\begin{equation}
\lim_{\alpha \rightarrow \infty}
{\left\{R^{^{\varphi}}_{T}(\alpha)\right\}}
=
\lim_{\alpha \rightarrow \infty}
{\left\{R_{T}(\alpha)\right\}}
=0
\label{518}
\end{equation}\normalsize
and
\small\begin{equation}
\lim_{\alpha \rightarrow 0}
{\left\{R_{T}(\alpha)\right\}}
= 1+ \frac{1}{2 n}, ~~~~
\lim_{\alpha \rightarrow 0}
{\left\{R^{^{\varphi}}_{T}(\alpha)\right\}}
= 1+ \frac{1}{n}
\label{519}
\end{equation}\normalsize

Both of them present the same asymptotic behavior which, at first glance, recover the theoretical possibility of a superluminal transmission in the sense that, by now, the SPM can be correctly applied since the analytical limitations are accurately observed.
At this point, it is convenient to notice that the superluminal phenomena, observed in the experiments with tunneling photons and evanescent electromagnetic waves \cite{Nim92,Ste93,Chi98,Hay01}, has generated a lot of discussions on relativistic causality.
In fact, superluminal group velocities in connection with quantum (and classical) tunnelings were predicted even on the basis of tunneling time definitions more general than the simple Wigner's phase time \cite{Wig55} (Olkhovsky {\em et al.}, for instance, discuss a simple way of understanding the problem \cite{Olk04}).
In a {\em causal} manner, it might consist in explaining the superluminal phenomena during tunneling as simply due to a {\em reshaping} of the pulse, with attenuation, as already attempted (at the classical limit) \cite{Gav84}, i. e. the later parts of an incoming pulse are preferentially attenuated, in such a way that the outcoming peak appears shifted towards earlier times even if it is nothing but a portion of the incident pulse forward tail \cite{Ste93,Lan89}.
In particular, we do not intend to extend on the delicate question whether superluminal group-velocities can sometimes imply superluminal signalling, a controversial subject which has been extensively explored in the literature about the tunneling effect (\cite{Olk04} and references therein).

Turning back to the scattering time analysis, we can observe an analogy between our results and the results interpreted from the Hartman Effect (HE) analysis \cite{Har62}.
The HE is related to the fact that for opaque potential barriers the mean tunneling time does not depend on the barrier width, so that for large barriers the effective tunneling-velocity can become arbitrarily large, where it was found that the tunneling phase time was independent of the barrier width.
It seems that the penetration time, needed to cross a portion of a barrier, in the case of a very long barrier starts to increase again after the plateau corresponding to infinite speed — proportionally to the distance \footnote{The validity of the HE was tested for all the other theoretical expressions proposed for the mean tunneling times \cite{Olk04}.}.
Our phase time dependence on the barrier width is similar to that which leads to Hartman interpretation as we can infer from Eqs.~(\ref{518}-\ref{519}).
Only when $\alpha$ tends to $0$ we have an explicit linear time-dependence on $L$ given by
\small\begin{equation}
t^{\varphi}_T = \frac{2\, m \, L}{w} \left(1 + \frac{1}{n}\right)
\label{14B}
\end{equation}\normalsize
which agree with calculations based on the simple phase time analysis where $t_T = \frac{2\, m \, L}{w} \left(1 + \frac{1}{2n}\right)$.
However, it is important to emphasize that the wave packets for which we compute the phase times illustrated in the Fig.(\ref{fig3A}) are not {\em effectively} constructed with the same momentum distributions.
The phase $\Theta(k, L)$ appears when we treat separately the momentum amplitudes $g(k - k_{\0})\,|T(k, L)|$ and $g(k - k_{\0})|R(k, L)|$ and the other one $\varphi(k, L)$ appears when we sum the amplitudes of the same side of the barrier.
In this sense, as a suggestive possibility for partially overcoming the incongruities (here pointed out and quantified) which appear when we adopt the SPM framework for obtaining tunneling phase times, we have claimed for the use of the multiple peak decomposition \cite{Ber04} technique presented in the study of the the above-barrier diffusion problem \cite{Ber04}.
We have essentially suggested a suitable way for comprehending the conservation of probabilities for a very particular tunneling configuration where the asymmetry presented in the previous case was eliminated, and the phase times could be accurately calculated.
An example for which, we believe, we have provided a simple but convincing resolution.

\section{Exact correspondence between phase times and dwell times}

\begin{figure}
\vspace{-0.6 cm}
\centerline{\psfig{file=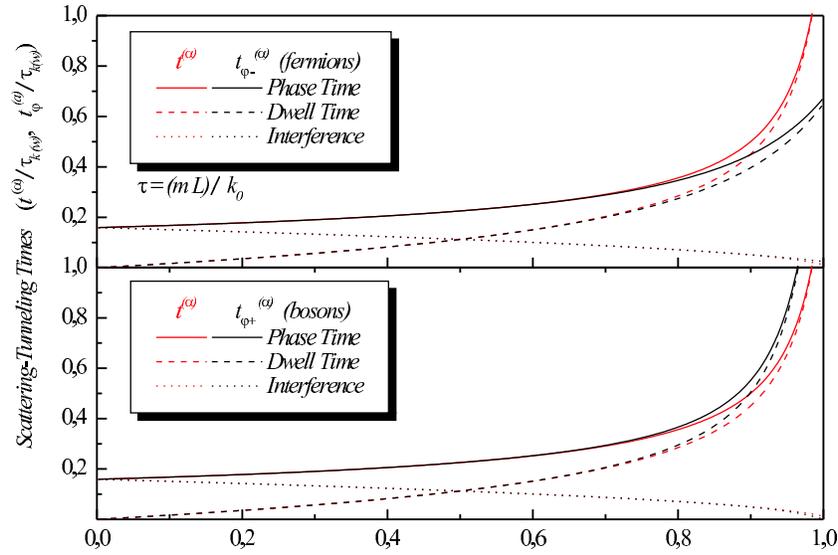,width= 11cm}}
\vspace{-1 cm}
\caption{Exact phase time (solid line), self-interference delay (dotted line), and the dwell time (dashed line) as a function of the normalized energy $n = k^{\2}/w^{\2} \propto E_{\0}/V_{\0}$ for the two identical particles (black line) and the standard one-way direction (red line) collision with a rectangular potential barrier.
For the colliding two identical fermions (first plot) we have assumed that the wave function is totally antisymmetrized.
For the colliding two identical bosons (second plot) we have assumed that the wave function is totally symmetrized.
These times are normalized by the {\em classical} traversal time $\tau_{\k} = (m L) /k$, and here we have adopted $w L = 4\pi$ for $\alpha = w L\sqrt{1-n}$.
\label{fig2}}
\vspace{-0.4 cm}
\end{figure}

At this point, one could say metaphorically that two bosonic particles represented by the symmetrized incident wave function spend a time equal to $ t_{T, \varphi_{\pl}}$ inside the barrier before retracing its steps or tunneling and that two fermionic particles represented by antisymmetrized incited wave function spend a time equal to $ t_{T, \varphi_{\mi}}$
The answer is in the definition of the dwell time for the same colliding configuration which we have proposed.
The dwell time is a measure of the time spent by a particle in the barrier region regardless of whether it is ultimately transmitted or reflected \cite{But83},
\small\begin{equation}
t_{D,\ppm}
=\frac{m}{k} \int_{\mi \L/\2}^{{\pl \L/\2}}\mbox{d}x{|\phi_{\pm,\2}(k,x)|^{\2}}
\label{530}
\end{equation}\normalsize
where  $j_{in}$ is the flux of incident particles and $\phi_{\2}(k,x)$ is the stationary state wave function depending on the colliding configuration that we are considering (symmetrical or standard).
To derive the relation between the dwell time and the phase time, we reproduce the variational theorem which yields the sensitivity of the wave function to variations in energy.
After some elementary manipulations of the Schr\"odinger equation \cite{Smi60}, we can write
\small\begin{equation}
\phi^{\dagger}\phi = \frac{1}{2m}\frac{\partial}{\partial x}\left(\frac{\partial \phi}{\partial E}\frac{\partial \phi^{\dagger}}{\partial x} - \phi^{\dagger}\frac{\partial^{\2}\phi}{\partial E\partial x}\right).
\label{531}
\end{equation}\normalsize
Upon integration over the length of the barrier we find
\small\begin{equation}
2 m \int_{\mi \L/\2}^{{\pl \L/\2}}\mbox{d}x{|\phi_{\2,\ppm}(k,x)|^{\2}} = \left.\left(\frac{\partial \phi}{\partial E}\frac{\partial \phi^{\dagger}}{\partial x} - \phi^{\dagger}\frac{\partial^{\2}\phi}{\partial E\partial x}\right)\right|_{\mi\L/\2}^{\pl\L/\2}.
\label{532}
\end{equation}\normalsize
In the barrier limits ($x = \pm L/2$), for the symmetrical configuration that we have proposed, we can use the superposition of the scattered waves to substitute in the right-hand side of the above equation,
\small\begin{eqnarray}
\left.\phi_{\ppm}(k,x)\right|_{\mi\L/\2(\pl\L/\2)} &=& \frac{\phi^{\L(\R)}_{\1}(k,x) \pm \phi^{\R(\L)}_{\3}(k,x)}{\sqrt{2}}
\nonumber\\
&=&
  \exp{\left[ \pm i \,k \,x\right]} + \exp{\left[ \mp i \,k \,x + i \left(\varphi_{\ppm}(k,L)- k L\right)\right]}
\label{533}
\end{eqnarray}\normalsize
By evaluating the right-hand side of the Eq.~(\ref{533}), we obtain
\small\begin{equation}
\frac{\partial k}{\partial E} \frac{d\varphi_{\ppm}}{dk} = \frac{m}{k} \int_{\mi \L/\2}^{{\pl \L/\2}}\mbox{d}x{|\phi_{\2,\ppm}(k,x)|^{\2}}
- \frac{Im[\exp{(i \varphi_{\ppm})}]}{k} \frac{\partial k}{\partial E}.
\label{534}
\end{equation}\normalsize
The first term of the above equation is the phase time or the aforementioned group delay $t^{(\alpha)}_{T, \varphi}$.
The second term leads to the explicit computation of the dwell time.
By respecting the continuity conditions of the Schr\"odinger equation solutions, in the barrier region we obtain a stationary wave symmetrical or antisymmetrical in $x$,
\small\begin{eqnarray}
\phi_{\2,\ppm}(k,x) &=& \frac{\phi^{\L}_{\2}(k,x) \pm \phi^{\R}_{\2}(k,x)}{\sqrt{2}}~~~~~~(\gamma\equiv\gamma^{\L,\R}\, \beta\equiv\beta^{\L,\R})
\nonumber\\
&=&
\sqrt{2}(\beta + \gamma)\frac{\exp{[\rho(k)\,x]} \pm \exp{[\rho(k)\,x]}}{2},
\label{534B}
\end{eqnarray}\normalsize
which, from Eq.~(\ref{530}), leads to
\small\begin{equation}
t^{(\alpha)}_{D, \varphi_{\ppm}} =
\frac{2\, \tau_{\k}\, n}{\alpha}
\frac{\alpha\pm\sinh{(\alpha)}}{2n - 1 \pm \cosh{(\alpha)}}
\label{535}
\end{equation}\normalsize
The self-interference term which comes from the momentary overlap between the incident and the reflected waves in front of the barrier is given by \footnote{We have printed the phase index $\varphi$ for all the results related to the symmetrical colliding configuration.}
\small\begin{eqnarray}
t^{(\alpha)}_{\I, \varphi_{\ppm}} &=& - \frac{Im[\exp{(i \varphi_{\ppm})}]}{k} \frac{\partial k}{\partial E}
=  \frac{m \, \sin{(\varphi_{\ppm})}}{k^{\2}} = \pm\frac{2 \tau_{\k}}{\alpha} \frac{(1-n)\sinh{(\alpha)}}{2n - 1 \pm \cosh{(\alpha)}}.
\label{536}
\end{eqnarray}\normalsize
It is interesting to observe that the result for the self-interference delay $t^{(\alpha)}_{\I, \varphi_{\ppm}}$ in the above equation also depends on the parity of the wave function.
The dwell time is obtained from a simple subtraction of the quote self-interference delay $t_{\I, \varphi_{\ppm}}$ from the phase time that describes the exact position of the peak of the scattered wave packets, $t_{\T,\varphi_{\ppm}} = t_{\D, \varphi_{\ppm}}+ t_{\I, \varphi_{\ppm}}$ as we can notice in the Fig.\ref{fig2}.

Adopting the {\em classical} traversal time $\tau_{\k} = (m L) /k$ for normalizing the results illustrated in the above  figures allows us to notice an important aspect concerned with the two identical fermion collision.
The two identical fermion collision leads to the possibility of an unusual {\em accelerated} tunneling transmission.

In fact, when each separated transmission coefficient $T^{\L,\R}$ prevails over the each reflection coefficient $R^{\L,\R}$, i. e. $|T|^{\2} > |R|^{\2}$, we have $|T|^{\2} > 1/2$.
For satisfying such a requirement the Eq.~(\ref{1}), after some manipulations, leads to $(w L)/(2\sqrt{n}) \leq w L \sinh{(\alpha)}) /(2\sqrt{n}) < 1 $.
For two identical bosonic particles, the possibility of accelerated tunneling transitions with respect to the traversal {\em classical} course is quantified by $t_{T, \varphi_{\pl}}^{(\alpha)}< \tau_{\k}$
It occurs only when $\alpha/2 \geq (\alpha/2) \tanh{(\alpha/2)} > 1$.
Since $\alpha = w L\sqrt{1-n}$, the intersection of the ``weak version'' of both of the above constraints, $(w L)/(2\sqrt{n}) < 1$ and $\alpha/2 > 1$, leads to $n > 2$, which definitely does not correspond to an effective tunneling configuration.
In the region where the one-way direction transmission coefficient prevails over the reflection coefficient, bosons should tunnel with a retarded velocity with respect to the classical velocity since, in this case, $t^{(\alpha)}_{T, \varphi}> \tau_{\k}$.
The supposed accelerated transit of the tunneling wave packet, and therefore, superluminal velocities and the Hartman effect, would never occur for $|T|^{\2} > |R|^{\2}$.
It does not correspond to the theoretical result for two fermionic particles since we can easily notice that $t_{\T, \varphi_{\mi}}^{(\alpha)}$ is always smaller than $\tau_{\k}$.
In fact, $t_{\T, \varphi_{\mi}}^{(\alpha)} < \tau_{\k}$ does not have a correspondence with the $R$ and $T$ coefficients.
Consequently, the possibility of accelerated tunneling and the verifiability of the Hartman effect for the  two identical fermion tunneling configuration is, in fact, concrete.

Claiming for the relevance of the use of the multiple peak decomposition technique, we have obtained the transit times for a symmetrized (two identical bosons) and an antisymmetrized (two identical fermions) quantum tunneling configuration.
We have demonstrated that the phase time and the dwell time give connected results in spite of the exact position of the scattered particles being explicitly given by the phase time (group delay).
For the antisymmetrized wave function configuration, an unusual effect of {\em accelerated} tunneling effect have been clearly identified in simultaneous two identical fermion tunneling.
In spite of quoting the superluminal interpretation, our discussion concerned with the definition of the strict mathematical conditions stringing the applicability of the stationary phase principle in deriving transit times.
Even with the introduced modifications, our results partially corroborate with the analysis \cite{Win03A,Win03} that gives an answer to the paradox of the Hartman interpretation \cite{Har62} since we provide a way of comprehending the conservation of probabilities \cite{Ber04, Ber06} for a very particular tunneling configuration where the asymmetry and the distortion aspects presented in the standard case were all eliminated.
Otherwise, one should keep in mind that {\em accelerated} tunneling transmission and, generically, the Hartman Effect, even in its more sophisticate consequences appears to have been experimentally verified \cite{Exp}, in particular, for opaque barriers and nonresonant tunneling \cite{Zai05}, and, under severe analytical restrictions, reproduced also by numerical simulations and constrained theoretical analysis \cite{Bar02,Pet03}.
In this sense, we suggest that the results here obtained deserve further attention by experimenter since the existing experiments, under the theoretical focus, report only about inaccurate results derived from non-symmetrical quantum tunneling constructions.

\section{Delay times for relativistic tunneling}

It is very difficult and probably even confusing to treat all interactions of plane waves or wave packets with a barrier potential using a relativistic wave equation \cite{Del03,Cal99,Dom99,Che02}.
This is because the physical content depends upon the relation between the barrier height $V_{\0}$ and the mass $m$ of the incoming (particle) wave, beside of its total energy $E$.
In the first attempt to evaluate this problem, Klein \cite{Kle29} considered the reflection and transmission of electrons of  incidence energy $E$ on the potential step $V\bb{x} = \Theta\bb{x}V_{\0}$ in the one-dimensional time-independent Dirac equation which can be represented in terms of the usual Pauli matrices \cite{Zub80} by \footnote{$\Theta\bb{x}$ is the Heavyside function.}
\small\begin{equation}
\left[\sigma^{3}\sigma^{\ii}\partial_{\ii} - (E - \Theta\bb{x}V_{\0}) - \sigma^{3} m\right]\phi\bb{k, x} = 0,
~~(\mbox{from this point}~ c = \hbar = 1),
\label{001}
\end{equation}\normalsize
which corresponds to the reduced representation of the usual Pauli-Dirac {\em gamma} matrix representation.
The physical essence of such a theoretical configuration lies in the prediction that fermions can pass through large repulsive potentials without exponential damping.
It corresponds to the so called {\em Klein tunneling} phenomenon \cite{Cal99} which follows accompanied by the production of a particle-antiparticle pair inside the potential barrier.
It is different from the usual tunneling effect since it occurs inside the energy zone of the Klein paradox \cite{Kle29,Zub80}.
Taking the quadratic form of the above equation for a generic scalar potential $V\bb{x}$,, we obtain the analogous Klein-Gordon equation,
\small\begin{equation}
\left(i \partial_{\0} - V\bb{x}\right)^{\2}\phi\bb{k, x} = \left(-\partial^{\2}_{\ii} + m^{\2}\right)\phi\bb{k, x},
\label{002}
\end{equation}\normalsize
which, from the mathematical point of view, due to the second-order spatial derivatives, has similar boundary conditions to those ones of the Schr\"odinger equation and leads to stationary wave solutions characterized by a {\em relativistically} modified dispersion relation.

By depicting three potential regions by means of a rectangular potential barrier $V\bb{x}$, $V\bb{x} = V_{\0}$ if $0 \leq x \leq L$, and $V\bb{x} = 0$ if $x < 0$ and $x > L$, differently from the non-relativistic (Schr\"odinger) dynamics, we observe that the incident energy can be divided into three zones.
The {\em above-barrier} energy zone, $E > V_{\0} + m$, involves diffusion phenomena of oscillatory waves (particles).
In the so called {\em Klein} zone \cite{Kle29,Cal99}, $E < V_{\0} - m$, we find oscillatory solutions (particles and antiparticles) in the barrier region.
In this case, antiparticles see an opposite electrostatic potential to that seen by the particles and hence they will see a well potential where the particles see a barrier \cite{Aux1,Kre04}.
The {\em tunneling} zone, $V_{\0} - m < E < V_{\0} + m$, for which only evanescent waves exist \cite{Kre01,Pet03} in the barrier region, is that of interest in this work.
In this context, our particular focus here is the calculation of phase times and dwell times and the possibility of a Hartman-like effect \cite{Har62}.

By evaluating the problem for this tunneling (evanescent) zone assuming that $\phi (k,x)$ are stationary wave solutions of the Eq.~(\ref{002}), when the peak of an incident (positive energy) wave packet reach the barrier $x = 0$ at $t = 0$, we can usually write
\small\begin{equation}
\phi(k,x)=
\left\{\begin{array}{l l l l}
\phi_{\1}(k,x) &=&
\exp{\left[ i \,k \,x\right]} + R(k,L)\exp{\left[ - i \,k \,x \right]}&~~~~x < 0,\nonumber\\
\phi_{\2}(k,x) &=& \alpha(k)\exp{\left[ - \rho\bb{k}  \,x\right]} + \beta(k)\exp{\left[ \rho\bb{k}  \,x\right]}&~~~~0 < x < L,\nonumber\\
\phi_{\3}(k,x) &=& T(k,L)\exp{ \left[i \,k (x - L)\right]}&~~~~x > L,
\end{array}\right.
\label{003}
\end{equation}\normalsize
where the dispersion relations are modified with respect to the usual non-relativistic ones: $k^{\2} = E^{\2} - m^{\2}$ and $ \rho\bb{k}^{\2} = m^{\2} - (E - V_{\0})^{\2}$.

The tunneling dynamics described by a relativistic wave equation can overcome all the (analytical) difficulties discussed in the last section.
We demonstrate with complete mathematical accuracy that, in some limiting cases of the relativistic (Klein-Gordon) tunneling phenomena where the relativistic kinetic energy is approximately equal to the potential energy of the barrier, and $m c L /\hbar << 1$, particles with mass $m$ can pass through a potential barrier $V_{\0}$ of width $L$ with transmission probability $T$ approximately equal to one.
Since $T \sim 1$, the analytical conditions for the stationary phase principle applicability which determines the tunneling (phase) time for the transmitted wave packets are totally recovered.
Differently from the previous (non-relativistic) tunneling analysis, the original momentum is kept undistorted and there is no {\em filter} effect.
We shall demonstrate that the tunneling time is then computed for a completely undistorted transmitted wave packet, which legitimizes any eventual accelerated transmission.

\subsection{Dynamics, variables and limits}

By assuming that the phase that characterizes the propagation varies smoothly around the maximum of $g(k - k_{\0})$, the stationary phase condition enables us to calculate the position of the peak of the wave packet (highest probability region to find the propagating particle).
With regard to the {\em standard} one-way direction wave packet tunneling, for the set of stationary wave solutions given by Eq.~(\ref{003}), it is well-known \cite{Ber06} that the transmitted amplitude $T\bb{n, L} = |T\bb{n, L}|\exp{[i \varphi\bb{n, L}]}$ is written in terms of
\small\begin{equation}
|T\bb{n, L}| = \left\{1 + \frac{1}{4 \, n^{\2} \, \rho^{\2}\bb{n}} \sinh^{\2}{\left[\rho\bb{n}\, w L \right]}\right\}^{-\frac{1}{2}},
\label{004}
\end{equation}\normalsize
where we have suppressed from the notation the dependence on $k$, and
\small\begin{equation}
\varphi\bb{n, L} = \arctan{\left\{\frac{n^{\2} - \rho^{\2}\bb{n}}
{2 n \, \rho\bb{n}}
\tanh{\left[\rho\bb{n} \, w L \right]}\right\}},
\label{005}
\end{equation}\normalsize
for which we have made explicit the dependence on the barrier length $L$ (parameter $w L$) and we have rewritten $\rho\bb{k} = w \rho\bb{n}$, with $\rho\bb{n}^{\2} = \sqrt{1 + 2 n^{\2} \upsilon} - (n^{\2} -\upsilon/2)$.

\begin{figure}
\epsfig{file=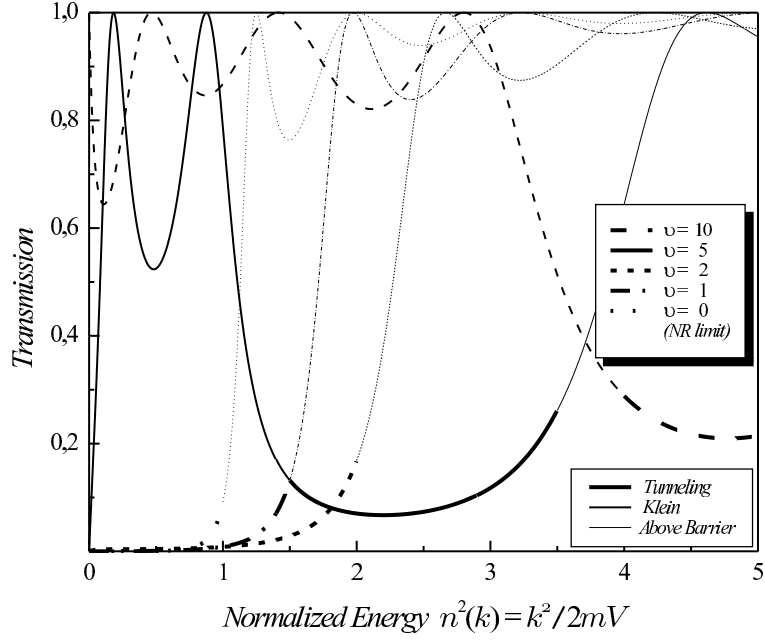,width=10cm}
\caption{Tunneling TRANSMISSION probabilities $|T\bb{n, L}|^{\2}$ for the incidence on a retangular potential barrier of height $V_{\0}$ for the dynamics of the relativistic wave equation versus the normalized energy $n^{\2}\bb{k}$.
We have classified the energy zones by the line thickness: the thick line corresponds to the tunneling zone, the intermediate line corresponds to the Klein zone and the thin line corresponds to the above-barrier zone.
Here we have adopted the illustratively convenient value of $w L = 2 \pi$ and have set $\upsilon =  0,\, 1,\, 2 ,\, 5,\, 10$, where NR regime can be parameterized by $\upsilon = 0$.
We have constrained our analysis to $n^{\2}\bb{k} > 0$ since we have assumed $V_{\0} > 0$.}
\label{Fig01}
\end{figure}

We illustrate the modified tunneling transmission probabilities in the Fig.(\ref{Fig01}) for different propagation regimes ($\upsilon = 0 (NR),\, 1,\, 2,\,5,\, 10)$) by observing that the tunneling region is comprised by the interval $(n - \upsilon/2)^{\2} < 1,\, n > 0$.

The additional phase $\varphi(n, L)$ that goes with the transmitted wave is utilized for calculating the transit time $t_{\varphi}$ of a transmitted wave packet when its peak emerges at $x = L$,
\small\begin{equation}
t_{\varphi} = \frac{\mbox{d}k}{\mbox{d}E\bb{k}} \frac{\mbox{d}n\bb{k}}{\mbox{d}k} \frac{\mbox{d}\varphi\bb{n, L}}{\mbox{d}n} = \frac{(L)}{v} \frac{1}{w (L)} \frac{\mbox{d}\varphi\bb{n, L}}{\mbox{d}n},
\label{006}
\end{equation}\normalsize
evaluated at $k = k_{\0}$ (the maximum of a generic symmetrical momentum distribution $g(k - k_{\0})$ that composes the {\em incident} wave packet).
By introducing the {\em classical} traversal time defined as $\tau_{\bb{k}} = L (\mbox{d}k/\mbox{d}E\bb{k})= L / v$, we can obtain the normalized phase time,
\small\begin{equation}
\frac{t_{\varphi}}{\tau_{\bb{k}}}
 = \frac{f\bb{n, L}}{g\bb{n, L}}
\label{007}
\end{equation}\normalsize
where
\small\begin{eqnarray}
f\bb{n, L} &=&
8 n^{\2} \left[\left(2 + 8 n^{\2} \upsilon + \upsilon^{\2}\right) - \left(4 n^{\2} + 3 \upsilon\right)\sqrt{1 + 2 n^{\2}\upsilon} \right]\nonumber\\
&&~~~~+
4 \left[\left(4 + 4 n^{\2}\upsilon + \upsilon^{\2}\right)\sqrt{1 + 2 n^{\2}\upsilon} - 2 \upsilon \left(2 + 3 n^{\2}\upsilon\right)\right]\frac{Sh(\rho\bb{n} w L)\,Ch(\rho\bb{n} w L)}{\rho\bb{n} w L}
\nonumber
\end{eqnarray}\normalsize
and
\small\begin{eqnarray}
g\bb{n, L} &=&
16 n^{\2} \left[2 \left(1 + 2 n^{\2}\upsilon\right) -  \sqrt{1 + 2 n^{\2}\upsilon}\left(2 n^{\2} + \upsilon\right)\right]\nonumber\\
&&~~~~+
2 \left[\left(4 + 8 n^{\2}\upsilon + \upsilon^{\2}\right)\sqrt{1 + 2 n^{\2}\upsilon} - 4 \upsilon \left(1 + 2 n^{\2}\upsilon\right)\right]Sh(\rho\bb{n} w L)^{\2}
\nonumber
\end{eqnarray}\normalsize
where $Ch(x) = \cosh{(x)}$ and $Sh(x) = \sinh{(x)}$.

\begin{figure}
\epsfig{file=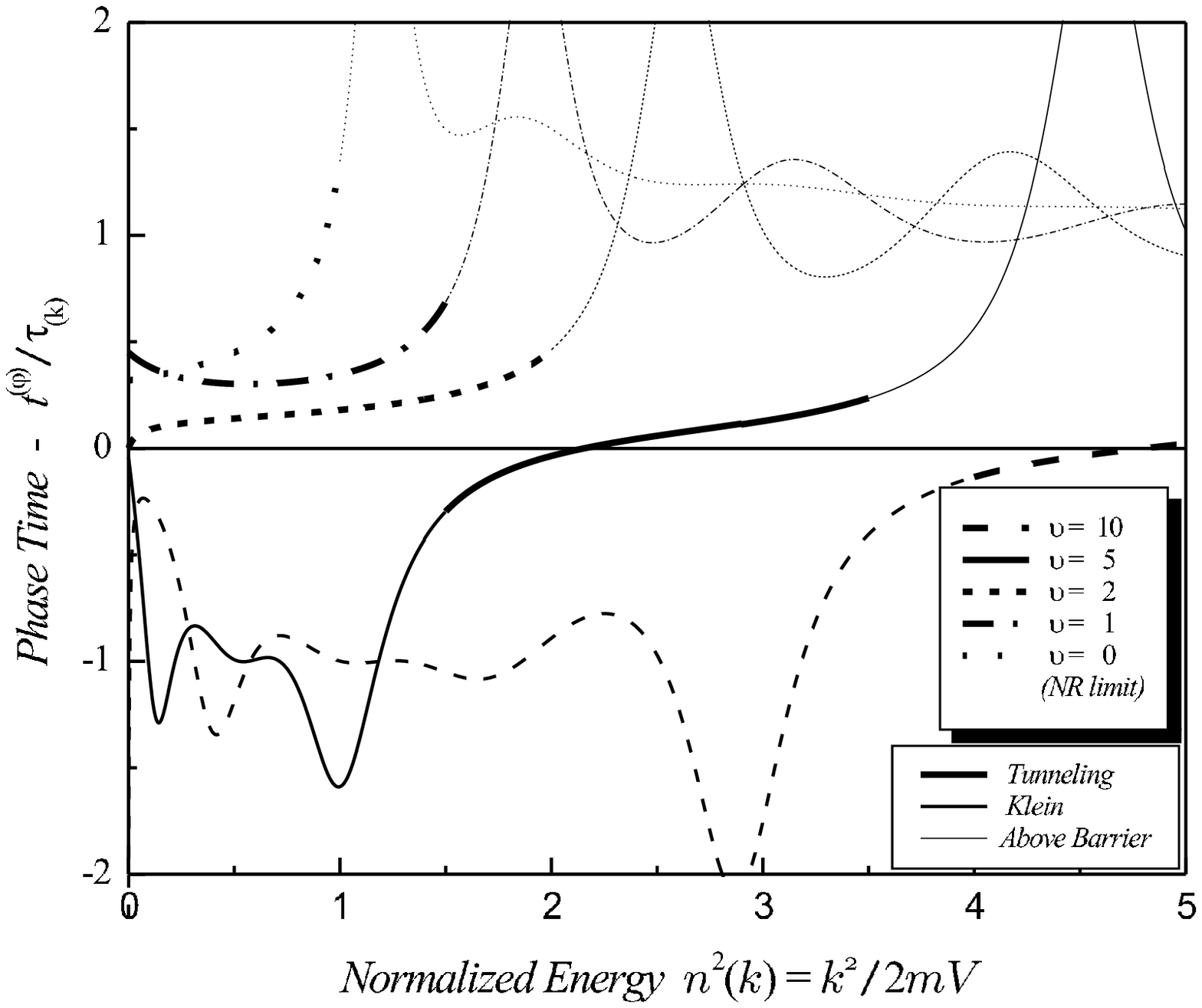,width=10cm}
\caption{The related tunneling PHASE times in correspondence with their the tunneling TRANSMISSION probability for the dynamics of the Klein-Gordon equation form.
Again by means of the line thickness we observe that the tunneling region is comprised by the interval $(n^{\2} - \upsilon/2)^{\2} < 1,\, n > 0$.
We have used the same criteria and the same set of parameters from the Fig.(\ref{Fig01}).}
\label{Fig02}
\end{figure}

To illustrate the above results we plot the tunneling phase times in correspondence with the
transmission probabilities of the Fig.(\ref{Fig02}) for the same different propagation regimes ($\upsilon = 0 (NR),\, 1,\, 2,\,5,\,10)$).

By observing the results illustrated by the above figures, we can notice the possibility of accelerated ($t_{\varphi} < \tau_{\bb{k}}$), and eventually superluminal (negative tunneling delays, $t_{\varphi} < 0$) transmissions without recurring to the usual analysis of the {\em opaque} limit ($\rho\bb{n} w L \rightarrow \infty$) which leads to the Hartman effect \cite{Har62}.
In the NR dynamics (Schr\"odinger equation solutions), the opaque limit and the consequent superluminal interpretation of the results of such an approximation (Hartman effect) were extensively discussed in the literature.
Superluminal group velocities in connection with quantum (and classical) tunneling were predicted even on the basis of tunneling time definitions more general than the simple Wigner's phase time \cite{Wig55} (Olkhovsky {\em et al.}, for instance, discuss a simple way of understanding the problem \cite{Olk04}).
Experiments with tunneling photons and evanescent electromagnetic waves \cite{Nim92,Ste93,Hay01} have generated a lot of discussions on relativistic causality, which, in addition to several analytical limitations, have ruined some possibilities of superluminal interpretation of the tunneling phenomena \cite{Lan89,Win03,Ber06}.
In a {\em causal} manner, the above arguments might consist in explaining the superluminal phenomena during tunneling as simply due to a {\em reshaping} of the pulse, with attenuation, as already attempted (at the classical limit) \cite{Gav84}, i. e. the later parts of an incoming pulse are preferentially attenuated, in such a way that the outcoming peak appears shifted towards earlier times even if it is nothing but a portion of the incident pulse forward tail \cite{Ste93,Lan89}.

The Hartman effect is related to the fact that for opaque potential barriers the mean tunneling time does not depend on the barrier width, so that for large barriers the effective tunneling-velocity can become arbitrarily large, where it was found that the tunneling phase time was independent of the barrier width.
It seems that the penetration time, needed to cross a portion of a barrier, in the case of a very long barrier starts to increase again after the plateau corresponding to infinite speed — proportionally to the distance \footnote{The validity of the HE was tested for all the other theoretical expressions proposed for the mean tunneling times \cite{Olk04}.}.

We do not intend to extend on the delicate question of whether superluminal group-velocities can sometimes imply superluminal signalling, a controversial subject which has been extensively explored in the literature (\cite{Olk04} and references therein).
Otherwise, the phase time calculation based on the relativistic dynamics introduced here offers distinct theoretical possibilities in a novel scenario, for the limit case where $\rho\bb{n}\,L$ tends to $0$ (with $L \neq 0$), in opposition to the opaque limit where $\rho\bb{n}$ tends to $\infty$.
Let us then separately expand the numerator $f\bb{n, L}$ and the denominator $g\bb{n, L}$ of the Eq.~(\ref{007}) in a power series of $\rho\bb{n}\,wL$ ($\rho\bb{n}  \rightarrow 0$) in order to observe that in the lower (upper) limit of the tunneling energy zone, where $n^{\2}$ tends to $\upsilon/2 + (-) 1$, the numerical coefficient of the zero order term in $\rho\bb{n}\,wL$ amazingly vanish in the numerator as well as in the denominator!
Since the coefficient of the linear term also is null, just the coefficient of the second order terms plays a relevant role in both series expansions.
After expanding the Eq.~(\ref{007}), such a {\em step-by-step} mathematical exercise leads to
\small\begin{equation}
\frac{t_{\varphi}}{\tau_{\bb{k}}}
 = \frac{4}{3}
\frac{\left[\left(4 + 4 n^{\2}\upsilon + \upsilon^{\2}\right)\sqrt{1 + 2 n^{\2}\upsilon} - 2 \upsilon \left(2 + 3 n^{\2}\upsilon\right)\right]}{\left[\left(4 + 8 n^{\2}\upsilon + \upsilon^{\2}\right)\sqrt{1 + 2 n^{\2}\upsilon} - 4 \upsilon \left(1 + 2 n^{\2}\upsilon\right)\right]}
+\mathcal{O}(\rho\bb{n}\,wL)^{\2}
\label{0100}
\end{equation}\normalsize
for small values of $\rho\bb{n}$.

\begin{figure}
\epsfig{file=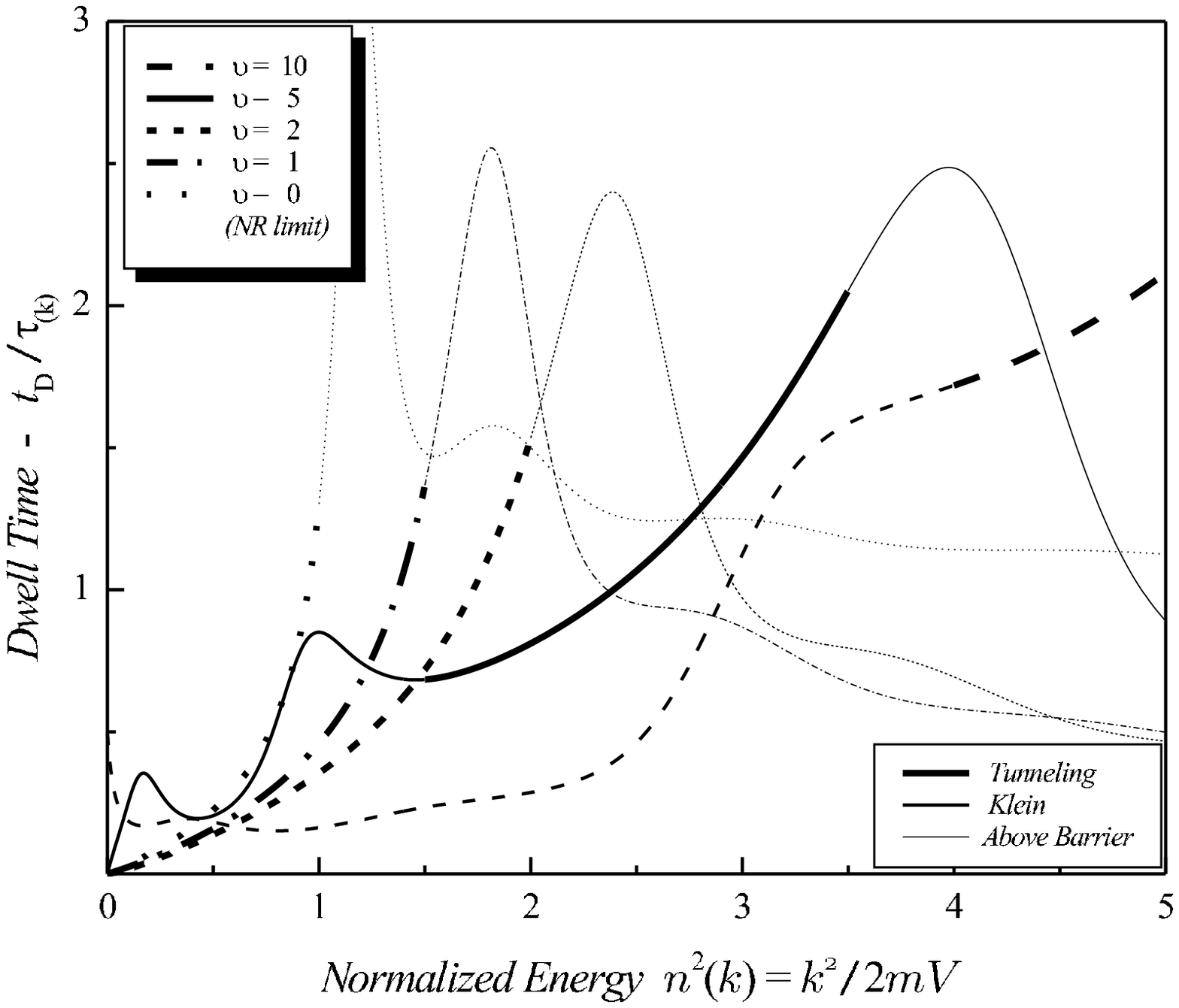,width=10cm}
\caption{The related tunneling PHASE times in correspondence with their tunneling TRANSMISSION probability for the dynamics of the Klein-Gordon equation form.
Again, by means of the line thickness we observe that the tunneling region is comprised by the interval $(n^{\2} - \upsilon/2)^{\2} < 1,\,\, n^{\2} > 0$.
We have used the same criteria and the same set of parameters from the Fig.(\ref{Fig01}).}
\label{Fig03}
\end{figure}

In parallel to the above way of calculating the traversal time, we believe that a more complete or, at least, complementary answer to such an inquiry is in the definition of the dwell time for the same tunneling configuration evaluated with a the Dirac equation in the Klein-Gordon equation form.
The dwell time is a measure of the time spent by a particle in the barrier region regardless of whether it is ultimately transmitted or reflected \cite{But83},
\small\begin{equation}
t_{\D}
=\frac{m}{k} \int_{\0}^{\L}{\mbox{d}x{|\phi_{\2}(k,x)|^{\2}}}
\label{008}
\end{equation}\normalsize
where  $j_{in}$ is the flux of positive energy incident particles and $\phi_{\2}(k,x)$ is the stationary state wave function inside the barrier.
The normalized dwell time is thus given by
\small\begin{equation}
\frac{t_{\D}}{\tau_{\bb{k}}} = \frac{f_{\D}\bb{n, L}}{g_{\D}\bb{n, L}}
\label{009}
\end{equation}\normalsize
where
\small\begin{eqnarray}
f_{\D}\bb{n, L} &=&\left(1 - \frac{n^{\2}}{\rho\bb{n}^{\2}}\right) + \left(1 + \frac{n^{\2}}{\rho\bb{n}^{\2}}\right) \frac{Sh(\rho\bb{n} w L)\,Ch(\rho\bb{n} w L)}{\rho\bb{n} w L},
\nonumber
\end{eqnarray}\normalsize
and
\small\begin{eqnarray}
g_{\D}\bb{n, L} &=& 2\sqrt{1 + 2 n^{\2}\upsilon} \left[1 + \frac{Sh(\rho\bb{n} w L)^{\2}}{4 n^{\2} \rho\bb{n}^{\2}}\right].
\nonumber
\end{eqnarray}\normalsize
We have computed the dwell times for the solutions of the relativistic wave equation (\ref{002}) and we have illustrated it in the Fig.(\ref{Fig03}) in correspondence with the transmission probabilities of the Fig.(\ref{Fig01}).

\begin{figure}
\epsfig{file=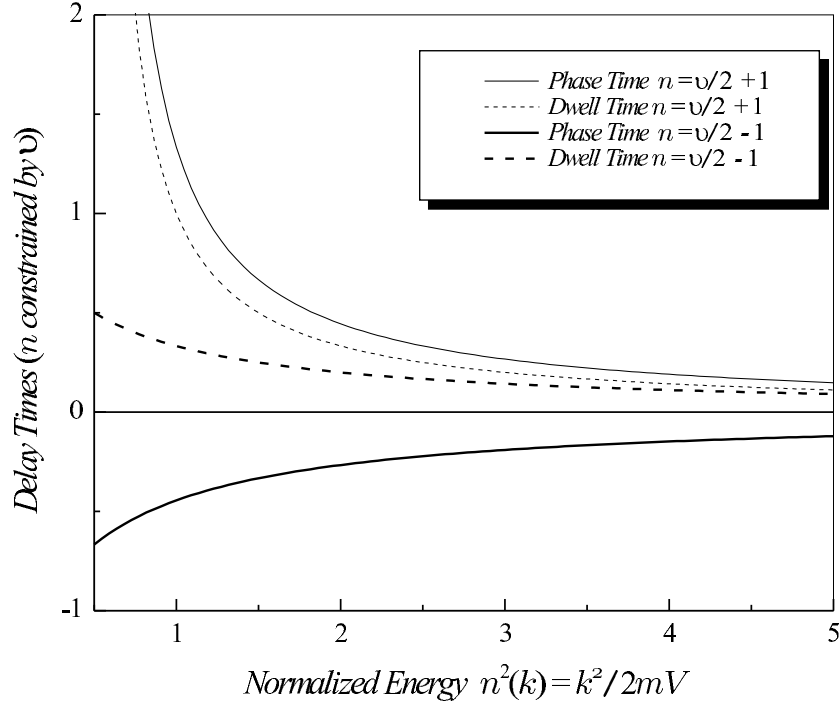,width=11cm}
\caption{Phase times (solid lines) and dwell times (dashed lines) in the limits of the tunneling energy zone as a function of the propagation regime (normalized energy parameter $n^{\2}\bb{k}$).
The analytical transition between the tunneling zone and the above-barrier zone is given by $n^{\2}= \upsilon/2 + 1$, and between the tunneling zone and the Klein zone by $n^{\2}= \upsilon/2 - 1$.
In the lower limit of the tunneling energy zone, $n^{\2}\rightarrow \upsilon/2 - 1$, it is always negative.}
\label{Fig04}
\end{figure}

By taking the same limits that we have considered for the phase time, the expressions for the dwell time provide us with,
\small\begin{equation}
\lim_{n^{\2}\rightarrow \upsilon/2 \mp 1}{\frac{t_{\D}}{\tau_{\bb{k}}}} = \frac{1}{2}\frac{1}{2 n^{\2} \pm 1}, ~~~~ n^{\2}\rightarrow \upsilon/2 \mp 1, ~~~n^{\2},\,\upsilon > 0.
\label{014}
\end{equation}\normalsize
from which we plot the comparative results in the Fig.(\ref{Fig04}).

The transmission probability depends only weakly on the barrier height, approaching the perfect transparency for very high barriers, in stark contrast to the conventional, non-relativistic tunneling where $T\bb{n, L}$ exponentially  decays with the increasing $V_{\0}$.
Such a relativistic effect is usually attributed that a sufficient strong potential which, being repulsive for electrons, is attractive for positrons, and results in positron states inside the barrier, which align the energy with the electron continuum outside \cite{Kat06}.

\subsection{Phase times and Dwell times - The relativistic case}

In order to reproduce the variational theorem for the Klein-Gordon equation (\ref{002}), we could write
\small\begin{equation}
\left(i \partial_{\0} - E \right)\phi\bb{k, x} = 0
\label{002A}
\end{equation}\normalsize,
its first derivative with respect to $E$,
\small\begin{equation}
\left(i \partial_{\0} - E \right)\frac{\partial \phi\bb{k, x}}{\partial E} - \phi\bb{k, x} = 0
\label{002B}
\end{equation}\normalsize
and, after some simple mathematical manipulations, its second derivative,
\small\begin{equation}
\left(\partial^{\2}_{\0} + E^{\2} \right)\frac{\partial \phi\bb{k, x}}{\partial E} + 2 E \phi\bb{k, x} = 0.
\label{002C}
\end{equation}\normalsize
By following the one-dimensional analysis here considered, it is easy to find that
\small\begin{eqnarray}
\left[\frac{\partial \phi}{\partial E} \frac{\partial^{\2}}{\partial x^{\2}}\phi^{\dagger} - \phi^{\dagger}\frac{\partial^{\2}}{\partial x^{\2}}\frac{\partial \phi}{\partial E}\right] &=& 2(E - V_{\0}) \phi^{\dagger}\phi\nonumber\\
&=& \frac{\partial}{\partial x}\left(\frac{\partial \phi}{\partial E}\frac{\partial \phi^{\dagger}}{\partial x} - \phi^{\dagger}\frac{\partial^{\2}\phi}{\partial E\partial x}\right).
\label{020}
\end{eqnarray}\normalsize
where we clearly notice the presence of $E - V_{\0}$ in place of $m$ which appears in the non-relativistic analysis \cite{Smi60} since in the strong non-relativistic limit $E - V_{\0} \mapsto m$.
Upon integration over the length of the barrier we find
\small\begin{equation}
2 (E - V_{\0}) \int_{\0}^{{\L}}\mbox{d}x{|\phi_{\2}(k,x)|^{\2}} = \left.\left(\frac{\partial \phi}{\partial E}\frac{\partial \phi^{\dagger}}{\partial x} - \phi^{\dagger}\frac{\partial^{\2}\phi}{\partial E\partial x}\right)\right|_{\0}^{\L}.
\label{021}
\end{equation}\normalsize
In front of the barrier ($x \leq 0$), the wave function consists of an incident and a reflected component given by $\phi_{\1}\bb{k, x}$, and behind the barrier ($x \leq L$), there is only the transmitted wave $\phi_{\3}\bb{k, x}$ (see Eq.~(\ref{003})).
Under these conditions we evaluate the right-hand side of Eq.~(\ref{021}) as
\small\begin{equation}
-2 i k \left[\frac{\mbox{d}}{\mbox{d}k}(|R|^{\2}+|T|^{\2}) + i \left((|R|^{\2}+|T|^{\2})\frac{\mbox{d}\varphi\bb{k, L}}{\mbox{d}k} + \frac{Im[R]}{k}\right)\right]\frac{\mbox{d}k}{\mbox{d}E}.
\label{022}
\end{equation}\normalsize
Since $|R|^{\2}+|T|^{\2} = 1$, Eq.~(\ref{021}) becomes
\small\begin{equation}
(E - V_{\0}) \int_{0}^{{\L}}\mbox{d}x{|\phi_{\2}(k,x)|^{\2}} =
\frac{\mbox{d}\varphi\bb{k, L}}{\mbox{d}E} + \frac{k}{E} Im[R]
\label{023}
\end{equation}\normalsize
which gives
\small\begin{equation}
\frac{t^{(\varphi)}}{\tau_{(k)}} = \frac{t^{(\D)}_{\R}}{\tau_{(k)}} - \frac{1}{\tau_{(k)}}\frac{Im[R]}{E}
\label{024}
\end{equation}\normalsize
where we have introduced the {\em re-scaled} dwell time,
\small\begin{equation}
t^{(\D)}_{\R}  = \frac{E - V_{\0}}{m} t^{(\D)} = \frac{E - V_{\0}}{k} \int_{\0}^{\L}{\mbox{d}x{|\phi_{\2}(k,x)|^{\2}}}
\label{025}
\end{equation}\normalsize
which can be related to the correct definition of the probability density for the Klein-Gordon equation,
\small\begin{equation}
j_{\0} = \int_{\0}^{\L}{\mbox{d}x\,[\phi^{\dagger}_{\2}(\partial_{\0}\phi_{\2}) - (\partial_{\0}\phi^{\dagger}_{\2})\phi_{\2}}]/ ~~~~ t^{(\D)}_{\R} = (j_{\0}/j_{in}) = (j_{\0}/ k),
\label{025B}
\end{equation}\normalsize
and leads to the usual definition $t^{(\D)}$ in the non-relativistic limit $(\frac{E - V_{\0}}{m} \mapsto 1)$.
In Eq.~(\ref{025}), the squared modulus of the wave function transforms as a Lorentz scalar,
$E - V_{\0}$ transforms as a time-like component, and the integrand $dx$ as well as $k$ transform as space-like components of a Lorentz four-vector.
It means that $t^{(\D)}_{\R}$ has the correct Lorentz character since it transforms as a time-like component, which does not occur for $t^{(\D)}$ of Eq.~(\ref{008}).

\begin{figure}
\epsfig{file=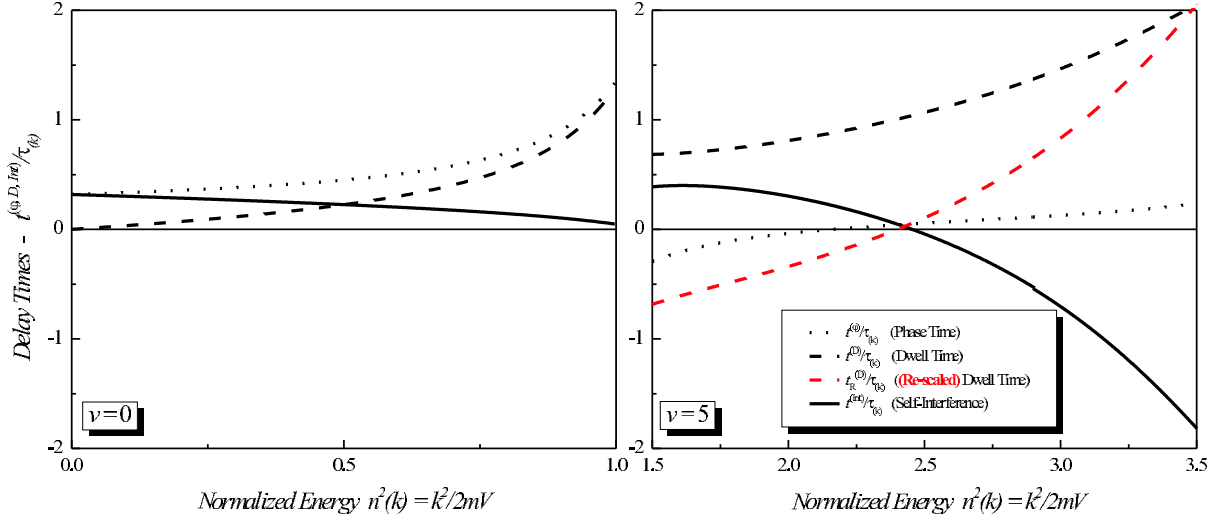, width=16cm}
\caption{Delay times calculated from the dynamics of the Klein-Gordon equation: Phase Time (Dash-dotted line), Dwell Time (Dashed black line), Self-Interference term (Solid line), and the {\em re-scaled} Dwell Time (Dashed red line).
In fact the tunneling region is comprised by the interval $(n^{\2} - \upsilon/2)^{\2} < 1,\, n^{\2} > 0$.
Here we have adopted the illustratively convenient value of $w L = 2 \pi$ with $\upsilon =  5$, in comparison with the non-relativistic results parameterized by $\upsilon \rightarrow 0$.
\label{Fig02B}.}
\end{figure}

As in the non-relativistic case \cite{Win03, Smi60}, the first term of Eq.~(\ref{024}) corresponds to the phase time or the aforementioned group delay.
The second term comes from the explicit computation of the dwell time.
However, the presence of the multiplicative factor $\frac{E - V_{\0}}{m}$ in Eq.~(\ref{025}) introduces some novel aspects in the interpretation of the additional term $-Im[R] / E$ as a self-interference term which comes from the momentary overlap of incident and reflected waves in front of the barrier \cite{Win03}.
As we can observe in the example illustrated in the Fig.~\ref{Fig02B}, and by the usual definition (\ref{008}), the dwell time is always positive.
The {\em re-scaled} dwell time modulated by $(\frac{E - V_{\0}}{m}$ changes sign when the total energy $E$ equalizes the potential energy $V_{\0}$: an energy region comprised by the tunneling energy zone of the Klein-Gordon equation.
Consequently, differently from the results we get from the non-relativistic analysis, the dwell time is not obtained from a simple subtraction of the {\em supposed} self-interference delay $t^{\bb{Int}}$ from the phase time that, in some circumstances \cite{Ber06,Ber07A} describes the exact position of the peak of the scattered wave packets.

At least for the moment, the above (relativistic) results do not necessarily demand for a confront with the (non-relativistic) predictions derived from the opaque limit analysis which results in the filter effect and the superluminal tunneling.
To clear up this assertion, it is convenient to recover the limiting configurations ($n^{\2}\rightarrow \upsilon/2 \pmp 1$) of some of our previous results \cite{Ber07A} for which the tunneling transmission probability (\ref{004}) can be approximated by
\small\begin{equation}
\lim_{n^{\2}\rightarrow \upsilon/2 \pmp 1}{|T\bb{n, L}|} =
\left[1 + \frac{(w L)^{\2}}{2 \upsilon \mp 4}\right]^{-\frac{1}{2}}
\mbox{$\begin{array}{c}\mbox{\tiny$\upsilon >> 1$}\\ \rightarrow \\~\end{array}$}
\left[1 + (m L)^{\2}\right]^{-\frac{1}{2}},
\label{011}
\end{equation}\normalsize
from which, avoiding any kind of filter effect, we recover the probability of complete tunneling transmission when $m L << 1$, once we have $|T\bb{n, L}| \approx 1$.
For the correspondent values of the phase times we obtain \cite{Ber07A},
\small\begin{equation}
\lim_{n^{\2}\rightarrow \upsilon/2 \mp 1}{\frac{t^{(\varphi)}}{\tau_{(k)}}} = -\frac{4}{3}\frac{1}{1 \pm 2 n^{\2}}, ~~~~ n^{\2}\rightarrow \upsilon/2 \mp 1, ~~~n^{\2},\,\upsilon > 0,
\label{012}
\end{equation}\normalsize
that does not depend on $m L$, and we notice that its asymptotic (ultrarelativistic) limit always converges to $0$.
Curiously, in the lower limit of the tunneling energy zone, $n^{\2}\rightarrow \upsilon/2 - 1$, it is always negative.
Since the result of Eq.~(\ref{012}) is exact, and we have accurately introduced the possibility of obtaining total transmission ({\em transparent barrier}), our result ratifies the possibility of accelerated transmission (positive time values), and consequently superluminal tunneling (negative time values), for relativistic particles when $m L$ is sufficiently smaller than 1 $(\Rightarrow T \approx 1)$.
By observing that the barrier height has to be chosen such that one remains in the tunneling regime,
it is notorious that the transmission probability depends only weakly on the barrier height, approaching the perfect transparency for very high barriers, in stark contrast to the conventional, non-relativistic tunneling where $T\bb{n, L}$ exponentially  decays with the increasing $V_{\0}$.
Obviously, the above results correspond to a theoretical prediction, in certain sense, not so far from the experimental realization.
The above condition should be naturally expected since we are simply assuming that the Compton wavelength ($\hbar/(m c)$) is much larger than the length $L$ of the potential barrier that, in this case, becomes {\em invisible} for the tunneling particle.
In general terms, the relativistic quantum mechanics establishes that if a wave packet is spread out over a distance $d >> 1/m$, the contribution of momenta $|p| \sim m >> 1/d$ is heavily suppressed, and the negative energy components of the wave packet solution are negligible; the one-particle theory is then consistent.
If we want to localize the wave packet in a region of space (wave packet width $d$) smaller than or of the same size as the Compton wavelenght, that is $d < 1/m$, the negative energy solutions (antiparticle states) start to play an appreciable role.
The condition $d < L < 1/m$ (where $d < L$ is not mandatory) imposed over a positive energy component of the incident wave packet in the relativistic tunneling configuration excite the negative energy modes (antiparticles) and, qualitatively, report us to the Klein paradox and the creation of particle-antiparticles pairs during the scattering process which might create the intrinsic (polarization) mechanisms for accelerated and/or non-causal particle teletransportation.

\section{Summary and outlook}

We have examined the meaning of the group delay in barrier tunneling for non-relativistic dynamics, extended to some relativistic dynamics.
Our analysis was constructed upon the wave packet multiple peak decomposition which emerges when we study the one-dimensional potential scattering/tunneling.
Concentrated on some incongruities with the application of the method of stationary phase for deriving phase times,
we have extended the analysis to the study of the relativistic dynamics of the tunneling phenomena.

In the first part of our study, we have employed the SPM for analyzing the above-barrier diffusion problem.
The main point we have elaborated on shows that the results of the SPM depend critically upon the manipulation of the amplitude prior to the application of the method.
The method is inherently ambiguous unless we know, by some other means, at least the number of separate peaks involved.
Essentially, what we have demonstrated is that the barrier results can be obtained by treating the barrier as a two-step process.
This procedure involves multiple reflections at each step and predict the existence of multiple (infinite) outgoing peaks.

In the second part of our analysis, we have attempt to the physical framework where the study of a tunneling process is embedded.
In the same analysis which generically involves analytically-continuous {\em Gaussian} pulses, holomorphic functions which do not have a well-defined front, we have quantified some aspects inherent to the quoted {\em filter effect}.
In this case, we have observed that the amplitude of the transmitted wave is essentially composed by the plane wave components of the front tail of the {\em incoming} wave packet which reaches the first barrier interface before the peak arrival.
As we have noticed, the {\em filter effect} by itself leads to ambiguities in the interpretation of the wave packet speed of propagation.
As an alternative possibility to partially overcome these incompatibilities, we have claimed for the use of the multiple peak decomposition technique.
Thus, in the same framework where we have treated the above-barrier diffusion problem, we have suggested a suitable way for comprehending the conservation of probabilities for a very particular tunneling configuration where the asymmetry presented in the previous case was eliminated, and the phase times could be accurately calculated.
An example for which, we believe, we have provided a simple but convincing resolution which may either reignite the tunneling time controversy or help settle the issue.

In more general lines, there have also been some trying of yielding complex time delays for tunneling analysis, ultimately due to a complex propagation constant.
In this framework, the supposition of superluminal features is considered artificial since the transmitted peak is not related causally to the corresponding incoming peak \cite{Lan94}.
In parallel to the most sensible candidate for tunneling times \cite{Hau89,Lan94}, a phase-space approach have been use to determine a semi-classical traversal time \cite{Xav97,Sok90,Sok91}.
This semi-classical method makes use of complex trajectories which, by its turn, enables the definition of real traversal times in the complexified phase space.
It is also commonly quoted in the context of testing different theories for temporal quantities such as arrival, dwell and delay times \cite{Hau89,Lan94} and the asymptotic behavior at long times \cite{Jak98,Bau01}.

As a not so ambitious but suggestive possibility to partially overcome some of the incongruities here noticed and quantified, which appear when we discuss tunneling phase times, we have claimed for the use of the multiple peak decomposition \cite{Ber04} technique developed for the above-barrier diffusion problem.
Thus we have introduced a suitable way for comprehending the conservation of probabilities for a very particular tunneling configuration where the asymmetry presented in the previous case was eliminated, and the phase times could be accurately calculated.
We let for a subsequent analysis the suggestive possibility of investigating the validity of our approach when confronting with the intriguing case of multiple opaque barriers \cite{Esp03}, in particular, in the case of non-resonant tunneling.
Still concerning with the future theoretical perspectives, the symmetrical colliding configuration also offers the possibility of exploring some applications involving soliton structures.
In summary, the above discussion reinforces the assertion that the investigation of wave propagation across a tunnel barrier has always been an intriguing subject which is wide open both from a theoretical and an experimental point of view.

We have concluded that the transmission probability depends only weakly on the barrier height, approaching the perfect transparency for very high barriers, in stark contrast to the conventional, non-relativistic tunneling where $T\bb{n, L}$ exponentially  decays with the increasing $V_{\0}$.
Such a relativistic effect is usually attributed to a sufficient strong potential that, being repulsive for electrons, is attractive for positrons and results in positron states inside the barrier, which align the energy with the electron continuum outside \cite{Kat06}.

Finally, in the last part of the manuscript, we have been concentrated on the accurate calculation of phase times and dwell times for the energy zone of tunneling governed by a relativistic (Dirac/Klein-Gordon) wave equation.
The well-known Klein paradox - unimpeded penetration of relativistic particles through high and wide potential barriers - is one of the most exotic and counter intuitive consequences of quantum electrodynamics (QED).
The phenomenon is discussed in many contexts in particle, nuclear and astrophysics but direct tests of the Klein paradox using elementary particles have so far proved impossible.
Here we have quantified the conditions for the occurrence of accelerated and, eventually, superluminal tunneling transmission probabilities in order to eliminate the called filter (Hartman) effect from the transmitted waves.
By eliminating the filter effect, the transmission probabilities approximates the unitary modulus (complete transmission through a {\em transparent} medium).
As a result, the calculation of the phase time could be accurately evaluated in order to give the {\em exact} position of the transmitted wave packet (or particles), for which, in the standard non-relativistic analysis, due to the distortion of the original momentum distribution, the position of the peak is shifted.

By considering the magnitude of the parameter $m L$ ($m c^{\2}/[\hbar (c/L)]$ in standard units) for an electron with mass $\sim 0.5\,MeV$, and observing that in natural units we have $0.2\, MeV\ pm \sim 1$, we conclude that it should be necessary a potential barrier of width $L << 1 pm$ to permit the observation of the quoted superluminal transmission.
By principle, its observation makes the effect relevant only for some exotic situations as, for instance, positron production around super-heavy nuclei ($Z \sim 170$) \cite{Gre85} or evaporation of black holes through generation of particle-antiparticle pairs near the event horizon \cite{Pag05}.
The superposition of several barriers in a nanoscopic scale could perhaps provide a truly observable effect.

In the most common sense, the above condition should be naturally expected since we are simply assuming that the Compton wavelength ($\hbar/(m c)$) is much larger than the length $L$ of the potential barrier that, in this limit situation, becomes {\em invisible} for the tunneling particle.
The relativistic quantum mechanics establishes that if a wave packet is spread out over a distance $d >> 1/m$, the contribution of momenta $|p| \sim m >> 1/d$ is heavily suppressed, and the negative energy components of the wave packet solution are negligible; the one-particle theory is then consistent.
However, if we want to localize the wave packet in a region of space (wave packet width $d$) smaller than or of the same size as the Compton wavelenght, that is $d < 1/m$, the negative energy solutions (positron states) start to play an appreciable role.
This qualitative arguments report us to the Klein paradox and the creation of particle-antiparticles pairs during the scattering process which might create the intrinsic (polarization) mechanisms for accelerated and/or non-causal fermion teletransportation.
The condition $d < L < 1/m$ imposed over a positive energy component of the incident wave packet in the relativistic tunneling configuration excite the negative energy modes (antiparticles) in the same way that the motion of electrons in a semi-conductor is concatenated with the motion of positively charged {\em holes}.

Turning back to the context of the nanoscopic scale structures, the most challenging possibility of observing similar effects occurs for massless (or effective mass) Dirac fermions in graphene structures.
Even though the linear spectrum of fermions in graphene implies zero rest mass, their cyclotron mass approaches to $ 10^{\mi\2} m_e$ \cite{Nov05},
which increases the superluminal tunneling scale to 1 angstrom.
In spite of the theoretical focus, the results here obtained apply to some configurations which should deserve further attention by experimenters in the study of the graphene structures where the dynamics of the electron is described by a relativistic-like dynamics.
For bi-layer structures, due to the chiral nature of their quasiparticles, quantum tunneling in these materials becomes highly anisotropic, qualitatively different from the case of normal, non-relativistic electrons.
In the last years, it has been speculated that, from the experimental point of view, the graphene provides an effective medium for mimicking relativistic quantum effects where, for instance, massless Dirac fermions allow a close realization of Klein's gedanken experiment whereas massive chiral fermions in bilayer graphene offer an interesting complementary system that elucidates the basic physics involved.
In conventional two-dimensional systems, strong enough disorder results in electronic states that are separated by barriers with exponentially small transmittance \cite{Kat06}.
In contrast, in single- and bi-layer graphene materials all potential barriers are relatively transparent ($T\bb{n, L} \approx 1$): the quasiparticles in graphene exhibit a linear dispersion relation $E = \hbar k v_{f}$
that corresponds to the pseudo-ultrarelativistic limit of our analysis for pseudo-massless particles traveling with their Fermi velocity $v_{f}$.
In this case there are pronounced transmission resonances where T approaches unity for some particular geometric configurations, which does not allow charge carriers to be confined by potential barriers that are smooth on atomic scale.
Some authors have demonstrated experimentally and theoretically, that the biased graphene bilayer is a tunable semiconductor where
the electronic gap can be controlled by the electric field effect reaching values as large as $0.3$ eV \cite{Cas06}.
These results can have important implications for the development of carbon-based electronics.

Once it has been observed that the physical essence of the Klein paradox lies in the prediction that particles can pass through large repulsive potentials without exponential damping \cite{Cal99,Dom99}, in this scenario, the Klein-paradox and the accelerated tunneling transmission associated with relativistic-like phenomena at nanoscopic scale can be tested experimentally using graphene devices.
To sum up, as future perspectives we intend to investigate the appearance of an {\em equivalent} smaller effective mass value $M_{eff} << m$ due the minimal coupling of the charged particle magnetic momentum with an external magnetic field, in particular, for the massless electron propagation in a single-layer graphene which could introduce some novel ingredients for quantifying these peculiarities of the relativistic tunneling effect.

\begin{acknowledgments}
The author thanks for the financial support from the Brazilian Agencies FAPESP (grant 08/50671-0) and CNPq (grant 300627/2007-6).
\end{acknowledgments}

\newpage
\normalsize

\newpage
\footnotesize
\begin{table}
\begin{minipage}{14cm}
\begin{center}
\caption{The values of $k$ numerically obtained by increasing the barrier extension $L$.
The values are calculated in terms of the wave packet width $a$ for different values of the potential barrier high expressed in terms of $w \, a$.
We have fixed the incoming momentum by setting $k_{\0} \, a$ equal to 1.}
\begin{tabular}{c|ccccccc}
\hline
\hline
~~~~~~~~~$w \, a$& 1.5 & 2.0 & 4.0 & 6.0 & 8.0 & 10 & 20 \\
$ L / a $ &&&&&&&\\
\hline\hline
0.00 & 1.0000& 1.0000& 1.0000& 1.0000& 1.0000& 1.0000& 1.0000\\
0.05 & 1.0062& 1.0188& 1.1777& 1.4156& 1.6238& 1.7726& 1.9834\\
0.10 & 1.0235& 1.0648& 1.3799& 1.6769& 1.8547& 1.9397& 2.0051\\
0.15 & 1.0489& 1.1223& 1.5349& 1.8251& 1.9505& 1.9937& 2.0133\\
0.20 & 1.0794& 1.1825& 1.6571& 1.9178& 2.0000& 2.0204& 2.0203\\
0.25 & 1.1129& 1.2420& 1.7575& 1.9813& 2.0317& 2.0390& 2.0272\\
0.30 & 1.1478& 1.3001& 1.8430& 2.0289& 2.0562& 2.0551& 2.0342\\
0.35 & 1.1836& 1.3565& 1.9185& 2.0679& 2.0779& 2.0704& 2.0413\\
0.40 & 1.2196& 1.4116& 1.9874& 2.1025& 2.0986& 2.0857& 2.0484\\
0.45 & 1.2558& 1.4657& 2.0524& 2.1350& 2.1191& 2.1012& 2.0556\\
0.50 & 1.2921& 1.5194& 2.1155& 2.1668& 2.1399& 2.1170& 2.0628\\
0.55 & 1.3285& 1.5729& 2.1785& 2.1988& 2.1611& 2.1331& 2.0701\\
0.60 & 1.3649& 1.6266& 2.2429& 2.2314& 2.1828& 2.1495& 2.0775\\
0.65 & 1.4015& 1.6809& 2.3101& 2.2651& 2.2051& 2.1663& 2.0850\\
0.70 & 1.4383& 1.7360& 2.3819& 2.3002& 2.2281& 2.1834& 2.0925\\
0.75 & 1.4751& 1.7920& 2.4599& 2.3367& 2.2518& 2.2009& 2.1001\\
0.80 &    $*\footnote{For the values of $L$ marked with $*$, we can demonstrate by means of Eqs.~(\ref{b}-\ref{c}) that the modulated momentum distribution has already been completely distorted and the maximum loses its meaning in the context of applicability of the method of stationary phase.}
            $& 1.8489& 2.5466& 2.3751& 2.2761& 2.2188& 2.1078\\
0.85 &    $*$& 1.9065& 2.6456& 2.4154& 2.3013& 2.2371& 2.1155\\
0.90 &    $*$& 1.9646& 2.7627& 2.4578& 2.3272& 2.2558& 2.1234\\
0.95 &    $*$&    $*$& 2.9091& 2.5028& 2.3540& 2.2750& 2.1313\\
1.00 &    $*$&    $*$& 3.1137& 2.5504& 2.3818& 2.2947& 2.1392\\
\hline\hline
\end{tabular}
\end{center}
\end{minipage}
\end{table}

\end{document}